\documentclass[twocolumn,draft,prb]{revtex4-1}

\usepackage{graphicx}
\usepackage{dcolumn}
\usepackage{bm}
\usepackage{subfigure}
\usepackage{dsfont}
\usepackage{xspace} 
\usepackage[USenglish]{babel}
\usepackage{amssymb,amsmath}
\usepackage{mathtools}
\usepackage{relsize}
\usepackage{lmodern}
\usepackage{slantsc}
\usepackage{scalefnt}
\usepackage{tikz}
\usetikzlibrary{arrows,calc,scopes}
\usetikzlibrary{shapes.geometric}
\usepackage{color}
\usepackage{array}
\usepackage{ntheorem}
\allowdisplaybreaks[1]

\newcommand{\be}{\begin{equation}}
\newcommand{\ee}{\end{equation}}
\newcommand{\bse}{\begin{subequations}}
\newcommand{\ese}{\end{subequations}}
\newcommand{\bpm}{\begin{pmatrix*}}
\newcommand{\epm}{\end{pmatrix*}}
\newcommand{\bmm}{\begin{matrix}}
\newcommand{\emm}{\end{matrix}}
\newcommand{\Z}{\mathbb{Z}}

\newcommand{\F}{\mathfrak{F}}
\newcommand{\ii}{\mathrm{i}}

\makeatletter
\newcommand*{\Relbarfill@}{\arrowfill@\Relbar\Relbar\Relbar}
\newcommand*{\xeq}[2][]{\ext@arrow 0055\Relbarfill@{#1}{#2}}
\newcommand{\Lcs}{\mathcal{L}_{CS}}
\newcommand{\pd}{\partial}
\newcommand{\dd}{\mathrm{d}}
\newcommand{\lv}{\mathbf{l}}
\newcommand{\dphib}{\mathbf{d\boldsymbol\phi}}
\newcommand{\Dphib}{\boldsymbol{\Delta\phi}}
\newcommand{\km}{$K$ matrix\xspace}
\newcommand{\kms}{$K$ matrices\xspace}
\newcommand{\x}{\times}
\makeatother

\newtheorem*{cgcondition}{Complete-Gapping Condition}

\newcommand\numberthis{\addtocounter{equation}{1}\tag{\theequation}}

\begin{document}


\title{A \km Construction of Symmetry Enriched Phases of Matter}

\author{Ling-Yan Hung}
\email{jhung@perimeterinstitute.ca}
\affiliation{Department of Physics, Harvard University, Cambridge MA 02138, USA}
\author{Yidun Wan}
\email{ywan@meso.t.u-tokyo.ac.jp}
\affiliation{Department of Applied Physics, Graduate School of Engineering, University of Tokyo,\\
Hongo 7-3-1, Bunkyo-ku, Tokyo 113-8656, Japan}

\date{\today}

\begin{abstract}
We construct in the \km formalism concrete examples of symmetry enriched topological phases, namely intrinsically topological phases with global symmetries. We focus on the Abelian and non-chiral topological phases and demonstrate by our examples how the interplay between the global symmetry and the fusion algebra of the anyons of a topologically ordered system determines the existence of gapless edge modes protected by the symmetry and that
a (quasi)-group structure can be defined among these phases. Our examples include phases that
display charge fractionalization and more exotic non-local anyon exchange under global symmetry that correspond to general group extensions of the global symmetry group.


\end{abstract}

\maketitle
\tableofcontents
\makeatletter
\let\toc@pre\relax
\let\toc@post\relax
\makeatother
\section{Introduction}\label{sec:Intro}
The understanding of phases of matter has come a long way beyond the Landau paradigm\cite{Wen1989}. Different phases of matter cannot be simply classified by Landau's symmetry breaking and a corresponding order parameter. In the case of phases involving short-range entanglement (SRE), it is now realized that for a given preserved global symmetry $G_s$, they could be subdivided into many different phases that cannot be connected by any local, unitary transformation without breaking the symmetry. These phases, called \emph{Symmetry Protected Topological}  (SPT) phases, turn out to be classified by group cohomology $H^2(G_s,U(1))$ in two spacetime dimensions and are believed  to be classified by $H^d(G_s,U(1))$ in $d$-dimensional spacetime\cite{Chen2011e}. An independent study based on \km construction that is particularly powerful in studying Abelian symmetry groups have confirmed many of the group cohomology classification, and moreover shed light on the edge excitations and transport properties of these phases\cite{Lu2012a,Levin2012a}. Things are more interesting when phases possess long range entanglement (LRE). Even without symmetry, they already show a very rich bulk structure, and so far only a partial classification of them is known. It is expected that when symmetry is incorporated, where such phases are often dubbed \emph{Symmetry Enriched Topological} (SET) phases, they would be further subdivided into different phases, and the allowed action of the symmetry group can be very exotic as it is already anticipated in earlier work on projective symmetry group where charge fractionalization is one common feature there\cite{Wen2002}.

More recently, there is renewed interest in systematically constructing and classifying these SET phases, notably in Ref\cite{Essin2012,Mesaros2011,Hung2012a,Wang2012}. Here, we would like to extend the methods in Ref\cite{Lu2012a,Levin2012a} to constructing SET phases in 2+1 dimensions. Our extension confirms many of the results in Ref\cite{Essin2012,Mesaros2011,Hung2012a}, particularly regarding the conditions of charge fractionalization and generalization to non-local symmetry transformations. Moreover, the \km analysis allows us to study the edge excitations in the presence of boundaries. A (quasi)-group structure among phases with the same global symmetry and fusion algebra emerges as we consider stacking them together, which does not appear to be directly related to group cohomology, although such a relation was found in the case of SPT phases\cite{Lu2012a}. We also generalize constructions in Ref\cite{Lu2012a} to include some non-Abelian symmetry groups. 

The \km construction is most powerful in dealing with phases whose anyons are governed by Abelian statistics. However, our study has inspired us of a more general way to construct and perhaps ultimately to classify symmetric phases with non-Abelian anyons. We will comment on the general idea towards the end of the paper.

Since our construction amalgamates and generalizes several ideas, we would like to begin our discussion with a general overview that puts together the various building blocks necessary for the current paper and clarify a few core concepts. 

In section \ref{sec:SETapproach} we will introduce our \km approach, based on the ideas developed in Ref\cite{Lu2012a,Levin2012a} how \km can be taken as the starting point for constructing SPT phases. Then we apply this approach in section
\ref{sec:Z2xZ2} to construct LRE phases with symmetries. We will dwell particularly on the symmetry
enriched $\Z_2$ gauge theory and the double semion model, studying their edge excitations and
a quasi-group structure that emerges between the phases. 
This is then generalized in section \ref{sec:ZMZN} 
to $\Z_M$ symmetry in phases with $\Z_N\times \Z_N$ and related fusion algebras.  

To explore more exotic group
actions of the symmetry group beyond charge fractionalization, we study some examples in \ref{sec:rot}
that involve anyon exchange based again on phases with $\Z_2\times \Z_2$
fusion algebra. More exotic and elaborate examples based on phases with fusion $\Z_4\x\Z_4$
is discussed in section \ref{sec:Z2Z2xZ2}.  We collect these ideas, and summarise
the unifying principles behind these examples in section \ref{sec:beyond}, where
we construct also new phases accommodating  discrete non-Abelian group actions,
explicitly the Dihedral groups. 

We compare our results with existing results in the literature in section \ref{sec:Comp},
and
then conclude our discussion in section \ref{sec:Disc} with open questions.
 
Appendices A and B collect some technicalities. Appendix C, however, provides the \km version of the ``duality'' relation between  a SPT phase and a topological phase, where the latter descends from gauging the global symmetry in the former. This relation was first proposed  and realized in the string-net formalism in Ref\cite{Levin2012} and then further discussed in Ref\cite{Hung2012,Mesaros2011,Hung2012a}.   

\section{Symmetry enriched phases in $2+1$ dimension: An Overview}\label{sec:Over}

The main focus of the current paper is to construct examples where topological phases -- namely
phases that possess LRE -- are endowed with global symmetries. The theme
has received much attention recently, for example in Ref\cite{Mesaros2011,Essin2012,Hung2012a}. Several principles
underlie these discussions and constructions, and we would like to summarize them, along with a conceptual account for our approach,  before moving on to
our explicit constructions that concretely realize these principles. 

Any discussion of symmetry can hardly avoid the introduction of groups. Since
a number of groups would be introduced in our discussion, we would
like to catalog them here and explain briefly the role each plays, for clarity and later convenience. 

One important feature of LRE-phases is the emergence of non-local \emph{deconfined} quasi-particles. 
In 2+1 dimensions for instance, quasi-particles (anyons) displaying Abelian or non-Abelian anyonic statistics furnish
such examples. While physical observables are characterized by local, bosonic excitations, anyons
are non-local and cannot be physically excited in complete isolation. Nevertheless, the phases often exhibit 
``deconfined'' limits,  in which it is possible to keep various anyons so far apart
that a lot of the operators can be considered as acting locally on individual anyons present. This is analogous to
the familiar situation in gauge theories, in which physical excitations are necessarily gauge invariant, even though
it is often useful to think of them as composites being made up of gauge charged particles, particularly
in a ``deconfined'' limit when the charged particles can be, to some extent, isolated. In fact, a lot of these
LRE phases can be conveniently described by gauge theories, such as the familiar case of $\Z_2$ spin-liquid, where
a $Z_2$ gauge symmetry effectively \emph{emerges} in the ``deconfined'' limit.  Moreover, while it is unclear
whether a complete classification of these LRE phases exist, and very likely, any such complete classification would
invoke the Mathematics of tensor categories\cite{Levin2004,WenTensorCat2004},  the framework of gauge theories
alone already encompasses a large class of LRE-phase\cite{Kitaev2003a,Kitaev2006,Levin2003,Levin2004,Hu2012a,Mesaros2011} 
including many of the well-known paradigmatic
examples such as the $\Z_2$ spin liquid. Most of the examples discussed in this paper are within the gauge theory
framework, and thus we will frequently refer to ``gauge groups'' $\mathcal{G}$ in this sense. These gauge theories
will be taken as the starting point on which we impose global symmetries. This starting point enables us to characterize or label an anyon-- each topological sector-- by its gauge charge and flux. A \textbf{flux} is labeled by a conjugacy class of $\mathcal{G}$, while the associated \textbf{charge} takes value in the irreducible (projective) representations of the centralizer subgroup of the flux in $\mathcal{G}$.
Each anyonic excitation for any given gauge group would fall into one of the three categories: {\bf pure charge},  {\bf pure flux}, and {\bf dyon}, which has both a flux label and the associated charge label. In the rest of the paper, we refer to different anyons using these terminologies where appropriate. Actually, as one will see, when a global symmetry is incorporated, another group that behaves essentially like a gauge group may appear, as we will explore shortly below.  This group however is generally different from $\mathcal{G}$.

As discussed in the previous paragraph, $2+1$ dimensions LRE phases generally bear anyonic, low-energy excitations, Abelian and/or non-Abelian. The interactions of these anyons are described by a set of fusion rules, in the sense that when viewed from sufficiently far away, various anyons relatively close together can be treated as a single lump. The lump behaves essentially as another anyon, now with a different \emph{topological} charge and/or flux that descend from those of the constituent anyons in the lump. These fusion rules generally form an algebra $\F$, a fusion algebra, which in the case of Abelian anyons, is in fact an Abelian group. As discussed in Ref\cite{Essin2012} and later generalized in Ref\cite{Hung2012a}, $\F$ plays a central role in determining the possible ways a global symmetry could act.  In this paper, like in most other discussions of symmetries, the global symmetries form a group $G_s$. Since symmetry acts reasonably locally in many cases, they can be understood as acting on individual anyons roughly independently. As emphasized above, however, anyons are not physical excitations and are thus not directly a physical observable; therefore, it is conceivable that the physical states must transform linearly under the global symmetry -- particularly that means they must transform trivially under the identity operator of the symmetry-- such a restriction can be lifted on individual anyons. The simplest possibility is that the anyons live in projective representation spaces of $G_s$, in which case the anyons are considered to have undergone \emph{charge fractionalization}. 

There are more exotic likelihoods, as demonstrated in Ref\cite{Lu2012a,Mesaros2011} and also
in some of our examples constructed in this paper in sections \ref{sec:rot} and \ref{sec:Z2Z2xZ2}, where exchange of anyons are involved
and such symmetry transformations are not strictly local as opposed to the case of fractionalization.
Nonetheless, in all these cases, the fusion algebra/group $\F$ constrains admissible ways the anyons can transform, by demanding that the aggregate transformations on any group of anyons that fuse to a physical bosonic excitation
must be reduced to those corresponding to a linear representation, such that the identity operator acts trivially. 
In other words, we are effectively ``modding out'' transformations on anyons when the aggregate transformation
of the group of anyons that fuse to a boson is trivial. These transformations that are \emph{modded} out
constitute a linear representation of a subgroup $N_g$ of $\F$. In this sense therefore, $N_g$ also
behaves very much like some kind of \emph{gauge group}, although it should not be confused with $\mathcal{G}$
introduced earlier. They are generally different. 

To concisely describe and thus classify these non-trivial transformations, we can introduce yet another group $G$.  In Ref \cite{Essin2012}, $G$ is the central extension of $G_s$ by $N_g$.
In that case, elements of $N_g$ necessarily commute with those of $G_s$.
This has been generalized to other group extensions, where $G_s$ is the quotient subgroup $G/N_g$,
where $N_g$ is the normal subgroup of $G$. Anyons transform as linear representations of $G$, and these group extensions provide the platform of classifying projective, and actually more general
non-linear representations of $G_s$ in which the anyons may fall into.
In this fashion, the group actions even of an Abelian $G_s$ do not necessarily commute, examples of which have been seen in Ref\cite{Mesaros2011} and will be shown in this paper. And more generally, the group $G$ is itself non-Abelian, and we obtain,  to our knowledge, the first
of such an example implementing non-Abelian group action in the \km construction, as discussed
in section \ref{sec:beyond}.

It should be noted that such classification of physically admissible non-trivial actions of global symmetries $G_s$ on any non-local excitations have appeared elsewhere.  Most notably, in fermionic symmetry protected topological (SPT) phases , which involve only short-range entanglement, fermions nevertheless can transform projectively under $G_s$ as long as any pair of them transform linearly. Framing it in the language developed above, the fusion group can be taken as $\F= \Z_2$ and the projective representations can be understood as group extension of $G_s$ by $\Z_2$. In bosonic SPT phases, since the underlying excitations are already physical bosons, there is no notion of charge fractionalization, and in our language, the fusion group can be thought as $\F=\Z_1$, the trivial group.

Before we close our discussions on Abelian phases, let us comment that the classification of symmetric LRE phases via the idea of group extensions does not \emph{a priori} inform us whether a given phase possesses non-trivial edge excitations in the presence of a boundary. Here, non-trivial edge excitations refer to the edge modes of the anyons that cannot be gapped out without breaking the symmetry and thus remain gapless as protected by the symmetry. The virtue of an explicit construction using \kms is that one is able to explore the fate of the edge states as much as classifying them. Despite transforming in highly non-linear representations under $G_s$, there is no guarantee that the edge behaves also non-trivially. We found examples in which even very exotic transformation rules, implying charge fractionalization and more, can lead to a gapped edge that respects the global symmetry. Among those phases that do possess non-trivial edges, which feature gapless excitations or spontaneousely broken global symmetry $G_s$, it appears that there is a notion of a (quasi)-group structure between them, when one considers stacking them together as in Ref\cite{Lu2012a}. This is a quasi-group also because the identity is not a single element but contains those phases that have fully gapped edge state without breaking $G_s$. This is discussed in section \ref{subsec:GpStr}. It is yet not completely clear whether such a group structure can always appear for any gauge theories, or that they  are related to group cohomology, as in the case of SPT phases\cite{Lu2012a}. 

In acknowledging the central and similar role $\F$ plays in constraining
admissible non-trivial $G_s$ representations in phases both short range and long range entangled, 
and also the quasi-group structure that ties together several phases that share the same fusion algebra $\F$,
we deem it convenient to refer generally to these phases as Symmetry Enriched Phases (SEP)
and label classes of them with the same symmetry group $G_s$ and fusion algebra $\F$ that are related by the group structure of $SEP(\F,G_s)$. This is  in fact a unified notion of phases with symmetry that also encompass SPT phases: Fermionic SPT phases are classified by $SEP(\Z_2,G_s)$, and bosonic SPT phases by $SEP(\Z_1,G_s).$ 

Finally, let us comment on the situation of non-Abelian anyons. In the above discussion, we have very much restricted our attention almost entirely to Abelian anyons, whose fusion $\F$ is an Abelian group. This is also the major focus of our paper, where we make heavy use of the \km construction, which is appropriate for Abelian anyons. But, the discussion here, and also the discussion of group extensions discussed in Ref\cite{Hung2012a} have pointed to a general way to constructing LRE phases with symmetries, if not completely classifying them. The idea is  very much like the case of Abelian anyons, where different phases can be thought of as different embedding of the fusion group $\F$ inside a larger group $G$. Quite generally, particularly in the framework provided in Ref\cite{Kitaev2003a,Kitaev2006,Levin2003,Levin2004,Hu2012a,Mesaros2011} describing large classes of LRE phases where the fusion $\F$ forms a representation ring of a \emph{quantum group,} which is an algebra $\mathfrak{U}$,  an LRE phase possessing symmetries can be thought of as embedding $\mathfrak{U}$ within a larger quantum group $\mathfrak{U}\mathcal{G}$. Analogous to the case of Abelian anyons, the quotient algebra is then taken as the global symmetry.  This framework provides a natural way in which anyons, which fall into irreducible representations of the larger algebra, can be decomposed as a direct sum of irreducible representations of $\mathfrak{U}$, which in turn dictates how anyons transform under the global symmetry given by the quotient algebra. The embedding of a smaller invariant subgroup in a larger one employs the same Mathematics as in symmetry breaking, in which a large (gauge) group is broken to its invariant subgroup. For non-Abelian topological phases, the relevant Mathematics would be that employed in Hopf symmetry  breaking, which has been discussed in Ref\cite{Bais2002,Bais2003,Bais2009a} in the context of anyon condensation. Many ideas can be directly applied here. We shall report a more detailed discussion elsewhere\cite{Hung}.an

\section{Symmetry enriched phases in $2+1$ dimension: The approach}\label{sec:SETapproach}
In this section, we shall elaborate on our approach for studying LRE Abelian phases with symmetry. We take the formalism known as \km plus Higgs terms. This formalism was used in Ref\cite{Lu2012a} for studying bosonic and fermionic SPT phases in $2+1$ dimension. We shall first briefly review the relevant pieces of this formalism then extend it to the case of LRE phases with symmetry.

\subsection{The $K$ matrix $+$ Higgs term formulation} \label{subsec:review}
It is believed that $2+1$ dimensional Abelian topological phases, including SRE phases and LRE Abelian, can be described in a unified fashion as effective Chern-Simons (CS) theories in the $K$-matrix formulation due to Wen \textit{et al}\cite{Wen1989a,Wen1990,Wen1990b,Wen1995}, whose generic Lagrangian density reads
\be\label{eq:KLagrangian}
\Lcs=-\frac{1}{4\pi}a^I_{\mu}K_{IJ}\pd_{\nu}a^J_{\rho}\epsilon^{\mu\nu\rho} -a^I_{\mu}j^{\mu}_I\cdots,
\ee 
where Einstein's summation rule is assumed. The internal indices $I,J=1,2,\dots,N$ label a set of $N$ internal $U(1)$ gauge fields $a^I_{\mu}$, where the greek letters are spacetime indices. The \km satisfies $K_{IJ}=K_{JI}\in\Z$. 
A generic quasiparticle, however, is a fusion of the fundamental ones and may be characterized by an integer vector $\lv=(l_1,l_2,\dots, l_N)^T$, carrying $l_I$ unit of $a^I_{\mu}$ charge. The self statistics of a quasiparticle and the mutual statistics of two different quasiparticles $a$ and $b$ are respectively given by 
\be\label{eq:stat}
\begin{aligned}
&\theta_a/\pi= (K^{-1})_{IJ}l_a^I l_a^J\\
&\theta_{ab}/\pi=2 (K^{-1})_{IJ}l^I_a l^J_b.
\end{aligned}
\ee
A physical quasiparticle is a boson, characterized by a vector $\lv_B$  that satisfies $\theta_B/\pi=0$ and $\theta_{Ba}/\pi=0\pmod{2\pi}$ with arbitrary quasiparticle $\lv_a$. The ground state degeneracy (GSD) of the system placed on a torus is given by 
\be
\mathrm{GSD}=|K|,
\ee
where as an abuse of notation, $K$ is the determinant of the \km. The Lagrangian  in Eq. \eqref{eq:KLagrangian} describes SRE phases if $|K|=1$ and LRE phases if $|K|>1$. In this paper, we concentrate on the latter case. Moreover, if $K$ has the same number of positive and negative eigenvalues, which also implies $\dim K\in 2\Z^+$, it describes a non-chiral topological order. 

In the absence of any symmetry, one can condense the bosons by adding to $\Lcs$ potential terms:
\be
\Lcs'=\Lcs+\sum_{\lv\in\text{bosons}}\bigg(C_{\lv}\prod_{I}b^{l_I}_I+\mathrm{h.c.}\bigg),
\ee  
where each $C_{\lv}$ is constant, and $b_I$ is the annihilation operator of the fundamental excitation of $a^I$ type, with $b_I^{\dag}=b_I^{-1}$. Each such term $C_{\lv}\prod_{I}b^{l_I}_I+\mathrm{h.c.}$ is often called a \textit{Higgs term}. Note that this addition does not affect any topological properties of the system described by $\Lcs$. The \km theory makes it handy to study the edge states if the system has a boundary. The effective action of the edge theory is given by
\be
S_E=S^0_E+S^1_E,
\ee
where
\be
S^0_E=\frac{1}{4\pi}\int\dd t\dd x\sum_{I,J}(K_{IJ}\pd_t\phi_I\pd_x\phi_J- V_{IJ}\pd_x\phi_I\pd_x\phi_J)
\ee   
corresponding to $\Lcs$ is the effective description of the gapless edge excitations, with $\phi_I$ the edge field associated with $a^I$ and $V_{IJ}$ a constant, positive definite matrix that determines the velocity of the edge excitations, and
\be\label{eq:edgeHiggsGen}
S^1_E=\sum_{\lv\in\text{bosons}}C'_{\lv}\int\dd t\dd x\cos(l^I\phi_I),
\ee
which corresponds to the bulk Higgs terms. Canonical quantization of $S^0_E$ yields the Kac-Moody algebra
\be\label{eq:edgeKacM}
[\partial_x\phi_I(x),\partial_y\phi_J(y)]=\ii 2\pi(K^{-1})_{IJ} \partial_x\delta(x-y).
\ee 

For simplicity, when referring to a quasiparticle $l^Ia_I$ or its edge mode $l^I\phi_I$, hereafter we will most often simply specify only the charge vector $\lv$.  
Besides, since we are mostly interested in the fate of the edge states, hereafter we shall refer to Eq. (\ref{eq:edgeHiggsGen}) or simply the cosine functions therein as our \textbf{Higgs terms}.
And we shall from now on focus on the edge modes $\boldsymbol\phi$ exclusively.

\subsection{Edge Gapping Conditions in LRE Phases with Symmetry}\label{subsec:approach}
To gap out a bosonic edge mode, one needs to condense it at certain classical expectation value; however, the uncertainty principle due to Eq. \eqref{eq:edgeKacM} may prevent one from doing so. Any two bosons labeled by vectors $\lv_a$ and $\lv_b$ must satisfy the following canonical commutation relation, implied by Eq. \eqref{eq:edgeKacM}, 
\be\label{eq:edgeBcomm}
[l_a^I\partial_x\phi_I(x),l_b^J\partial_y\phi_J(y)]=\ii 2\pi(K^{-1})_{IJ}l_a^Il_b^J \partial_x\delta(x-y).
\ee
 It is then clear that if boson $\lv_a$ can condense, the above commutator must vanish for $a=b$, i.e., $\lv_a^T K^{-1}\lv_a=0$. Such a boson is called \textbf{self-null}. Furthermore, 
two bosons $\lv_a$ and $\lv_b$ are only allowed to condense simultaneously
if they are both  self-null, as well as \textbf{mutual-null}, namely $\lv^T_aK^{-1}\lv_b$=0.

We now summarise the necessary and sufficient condition for  
a non-chiral LRE Abelian phase with symmetry to attain a symmetry preserving yet fully gapped edge state.

\begin{cgcondition}
Given a non-chiral, Abelian, LRE phase characterized by a \km satisfying $\dim K=N\in 2\Z^+$ and $|K|>1$, with a global symmetry group $G_s$, in order that the edge modes in the phase can be completely gapped without breaking $G_s$, there must exist  at least one \textbf{complete} set $\mathbf{B_I}$ of independent self and mutual null bosons, of which the Higgs terms are invariant under the action of $G_s$, namely, $\forall g\in G_s$,  
\begin{align*}\label{eq:gapCond2}
g:&\sum_{\lv_a\in\mathbf{B_I}}C_{\lv_a}\cos( l^I_a\phi_I+\alpha_{\lv_a})\\
&\mapsto \sum_{\lv_a\in\mathbf{B_I}}C_{\lv_a}\cos( l^I_a\rho_{IJ}(g)\phi^J +\alpha_{\lv_a})\numberthis\\
&=\sum_{\lv_a\in\mathbf{B_I}}C_{\lv_a}\cos( l^I_a\phi_I+\alpha_{\lv_a})
\end{align*} 
where each $\lv_a \in  \mathbf{B_I}$ appears at least once.   
$\alpha_{\lv_a}$ is some arbitrary angle allowed by the symmetry transformation, whereas $\rho_{IJ}(g)$ is an abstract notation of the representation of $g\in G_s$. The completeness of the set follows from two criteria: first, any boson that is not in $\mathbf{B_I}$ 
is not mutual-null with at least one member in $\mathbf{B_I}$; second, the
set must consist of at least $N/2$ linearly independent charge vectors. 

If this is the case, all edge boson modes can be pinned at a vacuum expectation value without breaking the symmetry, and we call the corresponding LRE phase with symmetry \textbf{edge-trivial}.
\end{cgcondition}
This condition should be self explaining. It is adapted to the case of LRE phases with symmetry from its counterpart in Ref\cite{Lu2012a} for the SPT phases, where one finds detailed reasoning for the condition. Two remarks are in order, however. For a \km describing a chiral LRE phase, it would be impossible to gap all the edge modes, as there would always be excessive left or right moving bosons. Also, for each given \km model there may be more than one complete set $\mathbf{B_I}$ which by definition cannot be mutual null with each other, and any single set that is completely gapped
is sufficient to gapping the edge. 
\subsection{Representations of the Symmetry}

Given any Lagrangian, one could look for the symmetries that leave
it invariant. In our case, the Lagrangian comprises the Chern-Simons terms
and the Higgs terms. However, we will work backwards in the program
of studying non-trivial phases with symmetries. We would start with the 
\km theory with a fixed $K$ matrix that has the correct degeneracy appropriate for the phase we 
are interested in (a topological phase with $|K| =N>1$), then exhaust all possible group action on the excitations for a given symmetry group $G_s$. An element of $G_s$ acts on the anyons depending on the representations they fall into, as indicated in Eq. \eqref{eq:gapCond2}. We would consider a very general scenario where the group actions rotate a dyonic state in
addition to attaching a phase to a charge or flux excitation, which can be implemented by the following pair
\be
\rho(g)=\{W^g,\dphib^g\},\;\forall g\in G_s,
\ee
where $W^g$ is some $GL(N,\Z)$ matrix and $\dphib^g$ a constant
$N$-component vector, called a \textit{shift vector}, such that
\begin{align*}
\phi_I \to \eta W^{IJ}\phi_J + d\phi_I, \\
K\to W^T K W = \eta K,\numberthis\label{eq:WpreserveK}
\end{align*}
where the second line follows from our requirement that $K$ is fixed, and $\eta=1$ for unitary actions and $-1$ for anti-unitary actions such as time-reversal symmetry. But we shall not consider any time-reversal in this paper and hence set $\eta\equiv 1$ here onward. As reasoned in Secion \ref{sec:Over}, this representation is in general a projective (or even more general non-linear) representation of  $G_s$. In particular, to be consistent with the fusion group of the anyons, the action of the identity $\mathbf{e}\in G_s$ can transform individual anyons but has to preserve the physical quasiparticles, namely the bosons, modulo $2\pi$, i.e.,. 
\be\label{eq:bosontrans}
\l_B^I \left((W^\mathbf{e})^{IJ} \phi_J + \dphib^\mathbf{e}_I\right) = l_B^I \phi_I \pmod{2\pi},
\ee
where $\lv_B$ labels the charge vector of a boson. Then we \emph{a posteriori}  look for Higgs terms invariant under the above action, and finally check for the presence of any remaining ungapped edge modes, or a gapped edge that dynamically breaks the symmetry. 

One can immediately infer from Eq. (\ref{eq:bosontrans}) that the action of $\{W^\mathbf{e},\dphib^\mathbf{e}\}$ does not depend on $G_s$; it is simply the set of transformations that is as exotic as possibly allowed by the fusion group. It is equivalent to a projective representation of the identity of $G_s$. Recalling the discussion in Section \ref{sec:Over}, when $G_s$ is present, $\{W^\mathbf{e},\dphib^\mathbf{e}\}$ furnishes a linear representation of an emergent gauge group $N_g$. We thus re-emphasize here that a projective representation $\{W^g,\dphib^g|g\in G_s\}$ can be interpreted as a linear representation of a total group $G$ that is an extension of $G_s$ by $N_g$, 
This will be played out explicitly in our examples.

Here is how one computes $\{W^g,\dphib^g\}$. Because Eq. \eqref{eq:bosontrans} must hold for any boson, one readily sees that $W^\mathbf{e}\equiv \mathds{1}$. Then, one needs to solve Eq. \eqref{eq:bosontrans} for $\dphib^\mathbf{e}$ with $W^\mathbf{e}=\mathds{1}$ plugged in. One would obtain a vector with a number of integer parameters determined by the fusion group. Each choice of the parameters renders the corresponding $\dphib^\mathbf{e}$ a generator of the emergent gauge group $N_g$. Each such generator then determines what $N_g$ is for this choice.

With $\{W^\mathbf{e},\dphib^\mathbf{e}\}$ in hand, one can then solve for $\{W^g,\dphib^g\}$ by solving a set of  equations,  which follow from the \textit{group compatibility conditions} of $G_s$. This is a list of independent group multiplication relations that completely specifies the group structure. For a simple example, if $G_s=\Z_2$, there is just one such compatibility condition, $g^2=\mathbf{e}$, $\forall g\in\Z_2$, which is then translated into the following equations:
\begin{align*}
&(W^g)^2=W^\mathbf{e}=\mathds{1}\\
&W^g(W^g\boldsymbol\phi+\dphib^g)+\dphib^g=W^\mathbf{e}\boldsymbol\phi +\dphib^\mathbf{e}\\
\Rightarrow\quad & W^g\dphib^g+\dphib^g=\dphib^\mathbf{e}.
\end{align*}  
More complicated $G_s$ may have more than one and more complicated group compatibility conditions. Among all solutions for $W^g$, one can only choose those satisfying Eq. \eqref{eq:WpreserveK}. The shift vector $\dphib^g$ contains the parameters of $\dphib^\mathbf{e}$ and its own parameters in general. If the parameters in $\dphib^\mathbf{e}$ are all switched off, it clearly implies that $G_s$ stays
unextended: $G=G_s$. 
Otherwise, one may need to work out the group compatibility conditions to determine precisely the type of group extension and the total group, for each choice of the parameters. Although $G_s$ is an Abelian group in this paper, the actions by the projective representation of two distinct elements on an anyon do not necessarily commute. This is particularly true when at least one of the actions is represented by a nontrivial $W^g$. This non-commutativity is ubiquitous in our examples, e.g., as explicitly discussed in Section \ref{subsec:Z2Z2SGZ2G}.

\subsubsection{Note of caution: residual gauge symmetry}
It is important to realize that the \km construction suffers
from a lot of redundancy. Different \km can describe identical phases, if they
are related by relabeling of the anyons, leading to, 
$K \to X^T K X$, for some $X \in GL(N,\Z)$.  As aforementioned, we
would be working with specific \km which is known to describe 
the topological phases we are interested in. Even with a fixed
\km, however, one is still haunted by the relabeling redundancy, because 
there are residual reparametrization $X$ which keeps a given \km
fixed. As a result, different global symmetry transformations
may not be uniquely defined. For those related by $X$ in fact describe
precisely the same phase. More precisely 
two sets of  transformations $\{W^{g_i},\dphib^{g_i}\}$
and $\{\tilde{W}^{g_i},\tilde{\dphib}^{g_i}\}$
describe the same physics if they are related by the following\cite{Lu2012a}
\begin{align*}\label{eq:gaugeWgDphig}
&K= X^TK X\\
&\tilde{W}^{g_i} = X^{-1}W^{g_i} X,\numberthis \\
& \tilde{\dphib}^{g_i} = X^{-1}(\dphib + {\bf \Delta\phi} -  W^{g_i} {\bf \Delta\phi}),
\end{align*}
${\bf \Delta\phi}$ is an arbitrary vector with integer components. These relations will help in locating the most convenient representative among equivalent solutions for $W^g$ and $\dphib^g$.


\section{$SEP(\Z_2\x \Z_2,\Z_2)$ Phases}\label{sec:Z2xZ2} 
In this section we construct our first example of symmetry enriched topological phases characterized by fusion group $\Z_2\x\Z_2$ and global symmetry $\Z_2$.
It is known that this fusion group is shared by two admittedly distinct models of topological order, the double semion model described by the \km $(\begin{smallmatrix}2 & 0\\ 0 & -2\end{smallmatrix})=2\sigma_z$, and the Kitaev's toric code model defined by the \km $(\begin{smallmatrix} 0 & 2\\ 2 & 2 \end{smallmatrix})=2\sigma_x$. We shall incorporate $\Z_2$ symmetry to these two models in order in the following two sub sections, then study the relations between these phases.  

\subsection{$SEP(\Z_2\x \Z_2,\Z_2)$ from the Semion Model}\label{subsec:setSem}
As aforementioned, the double semion model is defined by the \km $2\sigma_z$, which is invariant under the $GL(2,\Z)$ transformation $\pm \mathds{1}_2$, and $\pm\sigma_z$. The quasiparticle content of this model is determined by the self statistics in \eqref{eq:stat} and described as follows in terms of the vectors $\lv^T=(l_1,l_2)$.
\begin{description}
\item[Semions] \hfill 

The self statistics $(K^{-1})_{IJ}l^I l^J=\pm 1/2 \pmod 2$ demands $l_1=l_2+(2m+1),\, l_1,l_2,m\in \Z$. Hence, a generic semion is 
\[
\lv_S=(l,l+2m+1),\, l,m\in \Z.
\]
The elementary semions (of opposite chiralities) are thus $s_L=(1,0)^T$ and $s_R=(0,1)^T$.
\item[Bosons] \hfill

The self statistics $(K^{-1})_{IJ}l^I l^J=(l_1^2-l_2^2)/2=0 \pmod 2$ sets $l_1,l_2\in 2\Z$.
Thus, a generic boson takes the form
\[
\lv_B^T=(2m,2n),\, m,n\in\Z,
\] 
which is an authentic boson because it has trivial mutual statistics with an arbitrary quasiparticle $\lv^T=(l_1,l_2)$: $2(K^{-1})_{IJ}l_B^I l^J=2(ml_1-nl_2)=0\pmod 2$. The elementary bosons are $(2,0)$ and $(0,2)$.
\item[Bosonic bound states of semions]\hfill

There are also quasiparticles consisting of both elementary semions that have bosonic self statistics but nontrivial mutual statistics with the semions. These take the following general form.
\[
\lv_{bb}^T=(2m+1,2n+1),\, m,n\in\Z,
\] 
where the subscript "$bb$" stands for bosonic bound states.
\item[Sets of independent condensable bosons] \hfill

A condensable boson is a boson as defined above, with
the additional requirement that its self statistics is identically zero, and not only zero modulo $2\pi$.  A condensable boson therefore has to satisfy
\[
(K^{-1})_{IJ}l_B^I l_B^J = 2 (m^2-n^2) = 0,  \Rightarrow m = \pm n 
\]
Multiple condensable bosons can condense at the same time \emph{only if} their mutual statistics is also identically zero. In this case therefore, the two independent sets are given by $\{k (2,  2 )\}$ and $\{k (2, -2)\}$, for all $k\in \Z$.  
\end{description}

Here we remark that fermions are not in the quasiparticle spectrum of the double semion model because $l_1^2-l_2^2\neq 2\pmod 4,\, \forall l_1,l_2\in\Z$, which disallows fermionic self statistics. Having listed out the quasi-particle in the model, we can try to solve
for all possible (projective) representations that  are consistent with
the fusion properties of the quasi-particles when incorporating
a global symmetry. The idea is that the identity
element of any global symmetry must act trivially on each and every bosonic
particle, a condition already described in Eq. (\ref{eq:bosontrans}).
Yet this does not necessarily imply that it acts trivially on \emph{all} quasi-particle
excitations, although that is one obvious option. Therefore the first step we
take is to solve for all possible non-trivial ``identity transformation'' compatible
with Eq. (\ref{eq:bosontrans}), which we label by $\{W^\mathbf{e}, \dphib^\mathbf{e}\}$. 

For the semion model, the solution to Eq. (\ref{eq:bosontrans}) is given by
\be\label{eq:SGZ2iden}
W^\mathbf{e}=1, \qquad {\dphib^\mathbf{e}} = \pi (n_1,n_2)^T, \,\, n_1,n_2\in\{0,1\} .
\ee
This means that we have altogether four different options at our disposal. 
We can pick one to be the identity transformation for each independent symmetry
group we introduce for each phase. Each such choice gives rise to an emergent $N_g=\Z_2$ when a global symmetry is incorporated. As we will see however, some of these difference
choices could still potentially lead to the same phase.  
Next we have to solve for the rest of the symmetry transformations
for a given symmetry group. For simplicity, we will consider incorporating a
$\Z_2$ global symmetry here. This requires solving for the transformation
corresponding to the single generator of the group, which we label as $\{W^g, \dphib(g)\}$.
Since $g^2 = 1$, this transformation must satisfy
\be \label{eq:consistencyZ2}
\begin{aligned}
&W^g \dphib^g + \dphib^g =   \dphib^\mathbf{e} \pmod {2\pi}, \\
&{(W^g})^2 = W^\mathbf{e} = 1.
\end{aligned}
\ee
There are several sets of $\{W^g, \dphib(g)\}$ that satisfy the above.
They are listed as follows:
\be
\begin{aligned}\label{eq:semZ2sols}
\{W^g = \pm 1, \dphib=  \pi/2 (n_1,n_2)^T, \,\, n_1,n_2\in \{0,1\}\}\\
\{W^g= \pm\sigma_z,\dphib=  \pi/2 (n_1,n_2)^T, \,\, n_1,n_2\in \{0,1\} \}
\end{aligned}
\ee
At first sight there are many possibilities. However, we note that when $W^g = -1$,
the transformation $\dphib$ is not invariant under residual gauge transformation, and can
be entirely gauged away. Therefore this choice corresponds to the same phase as the semion model without symmetry. Also we note that the choice for $t_1, t_2$ has no effect on
the transformation of any bosons up to shifts of multiples of $2n\pi$. 
Therefore it has no bearing on the allowed Higgs terms, and therefore different choices
of which would lead only to the same phase. Therefore w.l.o.g we will consider
the representative case where they are chosen to be zero. 
When $W^g$ is not the identity, the group
action corresponds to
swapping quasi-particles or anyons around. This will be considered in a later section.
We will focus on $W^g =1$, and consider separate choices of $n_1, n_2$ in turn. 

\begin{description}
\item[Case Ia:$ \mathrm{S}10$]For later convenience, 
we label the phase for $\{n_1,n_2\}= \{1,0\}$ by $S10$, where $S$ stands for the semion model.
The invariant Higgs terms are given by
\be
S_E^{\mathrm{S}01} = \sum_{m\in \Z} C_m  \cos(4m (\phi_1 -\phi_2)).
\ee
As a result, the bosons with charge vector $2(2n+1)(1,-1)$ which is shifted by $\pi$
under $\dphib(g)$ would either acquire a vev due to the Higgs terms above and thus
break the $Z_2$ symmetry, or would remain gapless. 

\item[Case Ib:$ \mathrm{S}01$ ]The case $S01$ works very similarly since it as far as the Higgs
terms are concerned, it is a relabelling
of bosons by $\phi_1 \to -\phi_2, \phi_2\to -\phi_1$. The edge therefore remains
gapless.

\item[Case II: $\mathrm{S}11$]
The invariant Higgs terms are given by
\be
S_E^{S11} = \sum_{m\in \Z} C_m  \cos(2m (\phi_1 -\phi_2))
\ee
Clearly, in this case all mutually condensable bosons in the set $2m\{1,-1\}$
can all be simultaneously gapped. Therefore the $S11$ phase has trivial edge.
\end{description}

\subsection{$SEP(\Z_2\x\Z_2,\Z_2)$ from the $\Z_2$ Spin Liquid}\label{subsec:setZ2}
The $K$ matrix taken as our starting point here is given by $K= 2\sigma_x$.
Similar to the semion model we begin by listing all the quasi-particles:
\begin{description}
\item[Self-commuting ``boson''] These are excitations that have bosonic self statistics; however, they do not have trivial mutual statistics with all other excitations, which is why they are labeled as ``bosons'' in quotes.  Their charge vectors are $\lv^T= \{(m, n)|m,n\in \Z,m=n+1\pmod 2\}$.
Hence, the elementary ``bosons'' are $(1,0)$ and $(0,1)$.
 
\item[Fermionic bound states] \hfill

These are the set of particles that have fermionic self statistics $= \pi \pmod{2\pi}$.
The charge vectors are given by $\lv^T=\{2n+1,2m+1\}$, $n,m \in \Z$. The
elementary ``fermion'' is given by charge vector $(1,1)$.

\item[Bosons] These are \emph{true} bosons with $2n\pi$ 
mutual statistics with all quasi-particle exciations, and $2n\pi$ self statistics. 
Repeating precisely the same exercise as in the case of semions, we arrive at the set
of charge vectors
\be
\lv_B^T = (2n,2m), \qquad n,m\in \Z.
\ee
\item[Sets of independent condensable bosons] \hfill

Straightforwardly, the two independent sets of mutually commuting condensable bosons are
$\{(2n,0)\}$ and $\{(0,2n)\}$.
\end{description}

Similarly to the semion model we can solve for all possible
``identity transformation''. The solution is identical to that in Eq. (\ref{eq:SGZ2iden}).
Consider again imposing a global $Z_2$ symmetry on the $Z_2$ gauge theory, we
then solve for sets of transformation matrix $\{W^g, \dphib(g)\}$. The 
distinct solutions are 
\be
\begin{aligned}
&\{W^g=1, \dphib^T=\pi/2 (n_1,n_2),\,\, n_1,n_2 \in\{0,1\}\} \\
&\{W^g=\pm\sigma_x, \dphib^T=\pi/2 (n_1,n_2),\,\, n_1,n_2 \in\{0,1\}\},
\end{aligned}
\ee
where we have already hidden a possible $\pi(t_1,t_2)$ in $\dphib^T$ under a rug
for the same reason as before. Focusing on the solution on the top line $W^g= 1$,
the distinct phases are as follows:
\begin{description}
\item[Case I: T10]
We adopt similar labelling of the distinct phases, and ``T'' is an allusion to the Toric code model due to Kitaev, which is a popular solvable model realizing the $\Z_2$ gauge theory.
Here, the allowed Higgs terms are
\be
S_E^{\mathrm{T}10} = \sum_{m\in \Z} C_m  \cos(2m \phi_2 )
\ee
This clearly exhausts an entire set of mutually commuting condensable boson. 
Therefore the $\mathrm{T}10$ phase has trivial edge. The $\mathrm{T}01$ phase is obtained by a relabelling
$\phi_1 \leftrightarrow \phi_2$, and thus gives the same phase.
\item[Case II: T11]
Here the allowed Higgs terms (from a single mutually commuting set of bosons) are
\be
S_E^{\mathrm{T}11} = \sum_{m\in\Z}C_m \cos(4m \phi_{1}).
\ee
Bosons with charge vector $2(2n+1)(1,0)$ therefore either remain gapless or
breaks the symmetry. Therefore $\mathrm{T}11$ has non-trivial edge.
\end{description}

\subsection{A (quasi)-group structure between the phases}\label{subsec:GpStr}
Now we would like to discuss a (quasi)-group structure that emerges by
superposing the distinct phases with global $Z_2$ symmetry we have obtained
using the semion model and the $Z_2$ gauge theory as the starting point.

Our discussion of a group structure closely follows that in Ref\cite{Lu2012a}. The basic idea
there is that one can define a group product structure 
between two phases $A$ and $B$, within a class
of phases with a given symmetry, by stacking
one on top of the other. The combined phase would
generally allow for extra Higgs terms, gapping further edge modes. 
When a group structure is well defined, one could
show that the combined phase, described by  a new \km that is the direct sum of those
of the component phases, can be transformed after appropriate reparametrizations, into 
a direct sum of a  trivial SPT phase with a gapped edge, and another that is a memeber $C$
of the original class of phases with the given symmetry. This
allows one to identify a group product $A \oplus B = C$

There is a crucial difference between SPT phases and our LRE phases. 
In the case of $SPT$ phases $|K|=1$,  this is preserved as we superpose
phases. This allows one to naturally dump the SPT phase whose gap is trivially gapped 
after we stack the phases. This is no longer the case when we have $|K|>1$. Therefore the group structure
we are aiming for is not strictly a group. But consider the following situation: suppose we put two phases, $A$ and $B$ together, each with a non-trivial edge, and put them together exactly
as in the procedure described above.
Suppose also that there exists a relabeling of the bosons such that the new reparameterized $K$ matrix becomes again a direct sum of two topological phases with the Higgs terms now diagonalized in each component phase, and that at least one of which has entirely gapped edges, and the other,
called phase $C$, is recognizable as one of the phases we defined before superposing. Then there is indeed
some notion of a group structure where the group product of $A\x  B = C$, and that all
phases with trivial edge are treated as the identity element.
As we will find below, such a group structure indeed exists, but the group product only
closes if we are allowed to include both the $\mathrm{S}$ and $\mathrm{K}$ models in the group product.

We consider superposing different phases with non-trivial edges found above in turn:
\\
\\
\noindent$\boldsymbol{\mathrm{T}11 \x  \mathrm{T}11}$. Consider superposing $\mathrm{T}11$ phase with whose edge modes
are denoted $\{\phi^1_L,\phi^1_R\}$ and another $\mathrm{T}11$ with edge fields $\{\phi^2_L,\phi^2_R\}$.
The \km of the combined system is the direct sum of that of the constituent models.
In this case therefore it is given by $\mathcal{K}_{\mathrm{T}11\mathrm{T}11}= 2\sigma_x \oplus 2\sigma_x $
One allowed sets of Higgs terms within a chosen set of mutually condensable bosons  are given by
\be\label{eq:HiggsKK}
\begin{aligned}
&S_E^{\mathrm{T}11\x  \mathrm{T}11} = \sum_{m\in\Z} C^1_m \cos(2m(\phi^1_L-\phi^2_L)) + \\
& \sum_{m\in\Z} C^2_m \cos(2m(\phi^1_R+\phi^2_R))
\end{aligned}
\ee
One can check that this exhausts the entire set of mutually condensable bosons. 
The combined phase is left with a trivial edge. To display the group structure, we
now considering a relabelling of modes given by the following conjugation
$\mathcal{K} \to X^T\mathcal{K} X$ for some $SL(4,\Z)$ matrix $X$:  
\be
X = \bpm
1&0&0&0\\
0&1&0&-1\\
1&0&1&0\\
0&0&0&1
\epm
\ee  
This matrix $X$ leaves $\mathcal{K}_{\mathrm{T}11\mathrm{T}11}$ invariant. However
one can check that under this reparametrization where $\bf{\phi} \to X^{-1}\bf{\phi}$,
the group action $\dphib= \pi/2 (1,1,1,1)$ after
the transformation becomes
\be
X^{-1}\dphib=\pi/2 (1,2,0,1)^T 
\ee
The entry with value 2 in the transformation vector above
acts trivially on physical bosons, whose charge vectors consist only of components
divisible by 2. Therefore, $\mathrm{T}11\x  \mathrm{T}11$ is indeed the direct sum of two phases 
that we have already encountered previously:
\be
 \mathrm{T}11\x  \mathrm{T}11 = \mathrm{T}10 \oplus \mathrm{T}01 \sim 1,
\ee
and each of $\mathrm{T}10$ and $\mathrm{T}01$ has trivial edge, and so we replace them by ``1'', the identity element.
\\
\\
\noindent$\boldsymbol{\mathrm{S}10 \x  \mathrm{S}10}$.
Following the same logic as before, by superposing two $\mathrm{S}10$ phases,
we arrive at the model $\mathcal{K}_{\mathrm{S}10\mathrm{S}10} = 2\sigma_z \oplus 2 \sigma_z$.
The allowed Higgs terms are
\be
\begin{aligned}
S_E^{\mathrm{S}10\x  \mathrm{S}10} =& \sum_{m\in\Z} C^1_m \cos(2m(\phi^1_L-\phi^1_R-\phi^2_L-\phi^2_R))+\\
& \sum_{m\in\Z} C^2_m \cos(2m(\phi^1_L-\phi^1_R+\phi^2_L+\phi^2_R))
\end{aligned}
\ee 
Despite the appearance of two independent sets of Higgs terms, one can see that
there are further mutually condensable bosons that breaks the symmetry. They
are $\{2m (\phi^1_L-\phi^1_R)\} $ and $\{2m( \phi^2_L+\phi^2_R)\}$. Therefore $\mathrm{S}10\x  \mathrm{S}10$
contains non-trivial edge. To make contact with the original $2\x 2$ \kms,
we consider again a conjugation transformation of $\mathcal{K}_{S10S10}$ by $\tilde{X}$
\be
\tilde{X}= \bpm[r]
1& 0& 2& 0\\ 
0& 1&-1& 0\\
2& 1& 1& -1\\ 
-2&-1&-2&1
\epm
\ee  
This transformation \emph{does not} leave  $\mathcal{K}_{S10S10}$ invariant.
It is transformed upon conjugation into $\tilde{X}^T\mathcal{K}\tilde{X} =  2\sigma_z
\oplus 2 \sigma_x$. Correspondingly the group action $\dphib^T= \pi/2(1,0,1,0)$ becomes
\be
\tilde{X}^{-1}\dphib = \pi/2 (3,-1,-1,3)^T \sim  \pi/2 (1,1,1,1)^T,
\ee
where we use symbol $\sim$ to mean that $\tilde{X}^{-1}\dphib$, when
acting on physical bosons is indistinguishable from the final  transformation vector
on the right.  Therefore we conclude that
\be
\mathrm{S}10 \x  \mathrm{S}10 = \mathrm{S}11 \oplus \mathrm{T}11 \sim  \mathrm{T}11 \,
\ee
where $\mathrm{S}11$ as we recall has trivial edge and we define our group structure
that is only sensitive to the phase that has non-trivial edge states. 
\\
\\
\noindent$\boldsymbol{\mathrm{S}10\x  \mathrm{S}10\x  \mathrm{T}11\sim \mathrm{S}10\x  \mathrm{S}10\x  \mathrm{S}10\x  \mathrm{S}10}$.
From the above, we can immediately conclude that 
\be
\mathrm{S}10\x  \mathrm{S}10\x  \mathrm{T}11 \sim \mathrm{T}11\x  \mathrm{T}11 \sim 1,
\ee
and that
\be
\mathrm{S}10\x  \mathrm{S}10\x  \mathrm{S}10\x  \mathrm{S}10 \sim \mathrm{T}11 \x  \mathrm{T}11 \sim 1
\ee

\noindent$\boldsymbol{\mathrm{S}10\x  \mathrm{S}01}$.
One can easily check that the combined phases allows the following set of Higgs terms
\be
\begin{aligned}
S_E^{S10\x  S01} = \sum_{m\in\Z}& C^1_m\cos(2m (\phi^1_L -\phi^2_R))\\ + &C^2_m\cos(2m(\phi^1_R - \phi^2_L))
\end{aligned}
\ee
which exhausts all mutually condensable bosons, and thus have a trivial edge.

Therefore, one may be tempted to collect all the phases characterized
by the fusion group $\Z_2\x\Z_2$ with $\Z_2$ symmetry and arrange
them according to the emergent group structure
\be\label{eq:SEPZ2xZ2fake}
\begin{aligned}
&SEP(\Z_2\x\Z_2,\Z_2)\supset\Z_4=\\
&\{ 1\sim [T00, T10, S00, S11], S01,\mathrm{T}11\sim S01^2,S10\}.
\end{aligned}
\ee
Let us also clarify here that by taking the phases $[T00, T10, S00, S11]$ to
be the identity of the group structure, it is not to be understood as identifying these phases.
In fact as also emphasized in \cite{Yuan-MingLu2013} these phases cannot be
connected smoothly without a phase transition or breaking the symmetry.
We note that this collection of phases in $\Z_4$ do not include phases that involve non-local transformations of the anyons where $W^g=\pm\sigma_z$ in the double semion model and $W^g=\pm\sigma_x$ in the toric code model. As we shall see in Section \ref{subsec:return}, 
there are additional phases whose edge always remains gapless, and that stacking them
together never lead to a gapped edge.

\section{Generalization to $\Z_N$ gauge theories with global $\Z_M$ symmetries}\label{sec:ZMZN}
The discussion in the previous section over endowing the semion/$Z_2$ gauge theories
with a $\Z_2$ symmetry can be readily generalized to the case of taking some (generalized)
$\Z_N$ gauge theories and introducing $\Z_M$ symmetry. 

A (generalized) $\Z_N$ gauge theory can be described by a $K$ matrix
of the following form:
\be\label{eq:ZnKmatrix}
K_{[N,l]} = \bpm
0& N \\
N &2 l
\epm,
\ee
where $N,l \in \Z$, and $l \in \{0,1,\cdots N-1\}$. They are in one-to-one correspondence
with the Dijkgraaf-Witten lattice gauge theories, or equivalently the TQD models\cite{Hu2012a}

Given $K(N,l)$, one could readily obtain the general form of
the physical bosonic excitations in the model. They are given by charge vectors
 $\lv_B= (l_1,l_2)^T$ of the form
\be
l_1 = N m_1, \qquad l_2 = Nm_2 + 2m_1l,
\ee
where $m_1,m_2 \in \Z$.

There are two independent sets of condensable bosons
\be
\textrm{A}:=\{m (0, N)^T \}, \qquad
\textrm{B}:\{\frac{Nr}{\gcd(N,l)} (N,l)^T\}_{r \in \Z }.
\ee

One could solve for the set of $\{W^\mathbf{e}, \dphib^\mathbf{e}\}$. It is given by
\be
W^\mathbf{e}=1, \qquad {\dphib^\mathbf{e}} = \frac{2\pi}{N}\vspace{0.5em}\bpm
n_1 - \frac{2l n_2}{N}\\ 
n_2\epm,
\ee
for $n_i \in \{0,1,\cdots N-1\}$. From Ref\cite{Essin2012} it is asserted that the allowed projective representations of the symmetry group consistent with the fusion algebra always take values in the fusion algebra itself. For any $l$, the fusion algebra is additively generated by $\dphib^\mathbf{e}$ with all possible choices of $n_1$ and $n_2$ values, which can be straightforwardly derived as
\be\label{eq:fAlgebraZn}
\F=\left\{
\begin{array}{ll} 
\Z_N\x\Z_N,\quad & l=0\vspace{0.5em}\\
\Z_{kN}\x\Z_{N/k},\quad & l\neq0
\end{array}\right.,
\ee
where $k=N/\gcd(2l,N)$ for $N\in 2\Z$, and $k=N/\gcd(l,N)$ for $N\in 2\Z+1$.
On the other hand, since we are considering a single $G_s= \Z_2$ symmetry,
the ``gauge group'' $N_g$ that is involved in extending $G_s$ is additively ($\hspace{-0.7em}\mod 2\pi$) generated by a particular $\dphib^\mathbf{e}$ with a specific pair of $n_1$ and $n_2$ values, as shown in the following equation.  
\be\label{eq:NgZn}
N_g=\left\{
\begin{array}{ll} 
\Z_{N/x},\quad & ln_2=0\vspace{0.5em}\\
\Z_{yN},\quad & 2lan_2< N\\
\Z_{N/z},\quad & 2lan_2\geq N\\
\end{array}\right.,
\ee
where 
\begin{align*}
&x=\min[\gcd(n_1,N),\gcd(n_2,N)],\\
&y=N/\gcd(2ln_2,N),\\
&z=\min[\gcd(|n_1-2ln_2/N|,N),\gcd(n_2,N)].
\end{align*}

The corresponding transformation generated by the generator of
a $Z_M$ global symmetry takes the form:
\be
W^g=1, \dphib^g = \frac{2\pi}{M}\bpm 
t_1 + \frac{1}{N}(n_1- \frac{2ln_2}{N})\vspace{0.5em}\\ 
t_2 + \frac{n_2}{N}\epm,
\ee
where here we are still focusing on cases that do not involve rotation of anyons,
and $t_i \in \{0,1,\cdots M-1\}$.

To determine whether the edge is gapped in each of these cases,
we compute the transformation of bosons in each of the two complete
condensable sets of bosons. If either set can be completely gapped,
the edge is gapped, but otherwise remain gapless.

The transformation of the bosons in each set is given by $\dphib^g$
\be
\begin{aligned}
&l^I_A d\phi^g_I = 2\pi m \frac{(N t_2+ \frac{n_2}{N})}{M}\\
&l^I_B d\phi^g_I = 2\pi r \frac{N^2 t_1+ Nlt_2 + Nn_1-n_2 }{\gcd(N,l)M}
\end{aligned}
\ee

The edge would be gapped if either transformation vanishes modulo $2\pi$
with no further constraint on $m$ or $r$. 
In general it would require specifying $M$ and $N$
and also the set $\{t_i,n_i\}$ before one could determine if an
edge has been gapped.
Nevertheless let us illustrate in a few examples some representative cases.

\subsection{$M=2$, $N=$ odd}
For simplicity, let us begin with a very specific example.
In this case, we find that as soon as $Nt_2+k_2$ is even, set A is
completely gapped, and thus the edge is gapped.
Therefore we need only to consider what happens if $Nt_2+k_2$ is odd.
In that case, we have to determine if set B can be gapped. 
Let $\gcd(N,l)=: x$, such that $N= x a$ and $l = x b$, with
$a,b$ relatively prime,
the condition that all set B bosons
are gapped is then given by
\be
a(Nt_1+ n_1) + b(N t_2-n_2) =0 \pmod{2}.
\ee
Recall that both $a$ and $Nt_2-n_2$ are assumed odd here.
This suggests that if $b$ is even,  the edge is gapped when $Nt_1+n_1$
is even, and for $b$ odd, so should $Nt_1+n_1$.

\subsection{Special case: $M=N$}
In this special case, the above shift transformation acting on any one set of the condensable bosons take on particularly simple forms:
\be
\begin{aligned}
& \l_{\mathrm{A}}^I\dd\phi^g_I = 2\pi m (t_2 + \frac{n_2}{N}),\\
&\l_{\mathrm{B}}^I\dd\phi^g_I = \frac{2\pi r}{\gcd(N,l)} (N t_1 + l t_2 + n_1- l n_2/N),
\end{aligned}
\ee 
and a non-trivial edge is formed if neither of the two sets of condensable
bosons can be completely gapped out without breaking the global $\Z_N$ symmetry.
From the behaviour of the bosons in set A, it is immediately clear that  whenever 
\be\label{eq:gapA}
n_2=0 \pmod N,
\ee set $A$ is gapped, independently of the value of $t_i$ and $n_1$.
In fact one can check that the value of $t_i$ is immaterial in the transformation
of \emph{any} physical bosons. Therefore they do not parameterize distinct phases
and will be dropped from now on. 

Suppose that $\gcd(N,l)= x$, so that we can write $N= x a$ and $ l = x b$ for $a,b$ relatively prime. Then the gapping of 
set B modes requires that 
\be\label{eq:gapB}
n_1- b n_2 = 0 \pmod{N}.
\ee
Non-trivial edges therefore arise if Eqs. (\ref{eq:gapA}, \ref{eq:gapB}) are
not satisfied at the same time.
This leaves, for each $l$ a set of phases with non-trivial edges parameterized by 
\be\label{eq:edgeN}
\bpm n_1\vspace{0.5em}\\n_2\epm = 
\bpm (s_1+ b n_2)_{\hspace{-0.8em}\mod{N}}\vspace{0.5em}\\  n_2\epm, \; s_1,n_2\in \{1,\cdots N-1\}.
\ee

\subsection{A quasi-group structure}\label{subsec:ZnQuasiG}
For the general case where $N>2$, we find ourselves in a large 
network of phases, and it is by no means obvious that the simple (quasi)- group
structure that we find for $N=2,M=2$ that arises as we stack multiple phases
on top of one another should also arise here. 
Rather than giving a complete survey of the matter, which seems much more complicated,
we restrict our attention to the case $N=M=3$ and $l=0$ and demonstrate, in this
restricted scenario, that there is still a group structure existing between the phases.

When $l=0$, the phases with non-trivial edges are parameterized by different $\dphib^g$
as follows:
\be
\dphib^g= \frac{2\pi}{N^2}(n_1,n_2)^T, 
\ee
where $n_1,n_2\in\{0,1,2\}$.
We focus on phases with symmetry here, as that with $n_1=n_2=0$ is equivalent to the usual topological phase without symmetry. Given that when $n_1=0$ ($n_2=0$), the corresponding phase can by fully gapped by condensing with the variable $3m\phi_R$, we shall focus only on the phases labeled by nonzero $n_1$ and $n_2$:  there are, up to interchanging $n_1$ and $n_2$ by renaming of quasi-particles
three phases, given by $(n_1,n_2) = (1,1), (2,2), (1,2)$.

\subsubsection{Stacking (11) with (12) or (22) with (12)}
In these two scenarios, we find that the edges of the aggregate phase
can be completely gapped out. The Higgs term take the following form
\be
\begin{aligned}
S_E^{11\oplus 12} = \sum_{m\in\Z} C^1_m \cos(3m(\phi^1_L-\phi_L^2))  \\
+C^2_m \cos(3m \phi^1_R+ \phi^2_R).
\end{aligned}
\ee

The corresponding Higgs terms for $S_E^{22\oplus12}$ takes the same form
as the above, except that the signs of $\phi_L^2$ and $\phi_R^2$ are flipped. 

\subsubsection{Stacking up three phases of the same kind}
It is not hard to check however, that stacking two phases of
the same kind lead to non-trivial edges still. The next simplest option is to
stack up three phases of the same kind. Consider for example stacking up three
(11) phases. 
One can check that there is a complete set of Higgs terms that gap 
out the edge, given by
\begin{align*}
&S_E^{11\oplus 11\oplus 11} = \sum_{m_1,m_2,m_3 \in\Z} C_{m_1,m_2,m_3} \\
&\x \cos\bigg(3m_1(\phi^1_L+ \phi^2_L + \phi^3_L) \numberthis \\
&+ 3m_2(\phi^1_R - \phi^2_R) + 3m_3(\phi^2_R-\phi^3_R)\bigg)
\end{align*}
The same set of Higgs terms applies also to stacking three of the (22) phases or (12) phases.

\subsubsection{Stacking two (11) and a (22), or vice versa}
The above results already give us hints of a group structure. But to prove our point, we consider
also this case. 
It turns out that this is again completely gapped. And one choice of
the complete set of Higgs terms are given by
\begin{align*}
S_E^{11\oplus 22\oplus 11} =& \sum_{m_1,m_2,m_3 \in\Z} C_{m_1,m_2,m_3} \\
&\x \cos\bigg(3m_1(\phi^1_L+ \phi^2_L )  \numberthis\\
&+ 3m_2(\phi^2_L + \phi^3_L) + 3m_3(\phi^1_R-\phi^2_R+\phi^3_R)\bigg)
\end{align*}

Having looked at the stacking above, we can recognize that the
emerged group structure corresponds to a $\Z_3$, if we identity (11) and (22), which is justified
from the fact that stacking three layers of (11), or two layers of (11) with one layer of (22) both
lead to a gapped phase. The $(12)$ phase is the inverse of both $(11)$ and $(22)$, again
pointing to identifying (11) and (22) in this group structure. (We note that we have not defined carefully the procedure to preserve the ground state degeneracy $|K|$ as in the case of the $\Z_2$ gauge theory and doubled semion model. However, the similarity with the previous case
makes it sufficiently evident that it should work very similarly here. )

\section{$SEP$ phases involving the rotations of quasi-particles.}\label{sec:rot}
As already mentioned while we analysed the $Z_2$ gauge theory and semion model in detail, there are interesting choices of symmetry transformation involving a transformation matrix $W^g$ that is not the identity. Such a possibility was already explored in the \km construction of SPT phases without topological order\cite{Lu2012a}. When there is topological order, such transformations have a particularly vivid physical interpretation.

\subsection{A return to $\Z_2$ theories with $\Z_2$ symmetry}\label{subsec:return}
Let us return to the $\Z_2$ gauge theory with \km $2\sigma_x$, and recall that the allowed choice of $W^g = \sigma_x$ which implements a global $Z_2$ symmetry on the theory. Its action on the $\phi$ is accordingly $\phi \to W^{g\,\,-1}\phi$, which alternatively, acts on the charge vector $\lv$ as $\lv \to W^{g\,T}\lv$. Recall that $\lv^T= (10)$ corresponds to the ``electric'' excitation, and that $\lv^T=(01)$ the ``magnetic'' excitation, this suggests that the action of $W^g=\sigma_x$ is precisely to exchange the anyons, implementing an \emph{electric-magnetic duality} in this case. In fact, more generally, whenever $W^g \neq 1$ it permutes the anyon excitation. Such a symmetry operation is non-local and is not considered in Ref\cite{Essin2012}.

We note also that whenever $W^g \neq 1$, only eigenvectors of $W^g$ could stay invariant, up to a sign (since determinant of $W^g= \pm 1$). However, since $W^g$ is directly proportional to the \km (and its inverse) itself, it implies immediately that these eigenvectors cannot be self-null at the same time. In other words, no condensable bosons could be left invariant by $W^g$ the $\Z_2$ gauge theory or the semion model. Therefore all of these phases have non-trivial edges, and no amount of stacking among these phases can lead to a gapped edge. In this case, $W^g$ can be either $\sigma_x$ or $-\sigma_x$, indicating a $\Z\x\Z$ group of phases with gapless states Note that we have relaxed our definition of a group structure here
compared to our discussion in Section \ref{subsec:GpStr} where a $\Z$ class is referring to the fact that we keep getting new phases as we stack phases
on top  without ever hitting a phase with trivial edge, although without defining a corresponding procedure to remove part of the system to preserve the torus ground state degeneracy $|K|$,
strictly speaking these extra phases may not belong to $SEP(\Z_2\x\Z_2,\Z_2)$ . 

The same consideration applies equally to the semion model, except that an admissible choice of $W^g$ which keeps  its \km invariant is given by $\pm\sigma_z$, indicating also $\Z\x\Z$ group of symmetry enriched phases. 


\subsection{More exotic examples: $\Z_2\x  \Z_2$ symmetries in $\Z_2$ gauge theories}\label{subsec:Z2Z2SGZ2G}
Such a global symmetry is considered also in Ref\cite{Mesaros2011}. When the symmetry group is a direct  product of groups, one could imagine that there are several relations among the groups. In the case of $\Z_2\x  \Z_2$, it amounts to the following:
\be
\begin{aligned}
& g_L^2 = 1,\quad g_R^2 = 1,\\
& g^{}_L g^{}_Rg^{}_Lg^{}_R = 1.
\end{aligned}
\ee
where $g^{}_L$ and $g^{}_R$ are respectively in the left and right $\Z_2$ factors of the global symmetry. For each relation, one needs not have the same choice of $\{W^\mathbf{e},\dphib^\mathbf{e}\}$ replacing the action of the identity, up to some consistency constraints. Had we chosen, however $W^{g^{}_L}=W^{g^{}_R}=1$ as in the previous sections, the only transformation has to come from the shifts $\dphib$. We would end up with the statement that the operators implementing $g^{}_L$ and $g^{}_R$ necessarily commute. And the analysis that follows from taking $W^{g^{}_L}=W^{g^{}_R}=1$ would be very much similar to what we have already considered in the previous sections,
which we will not repeat here.

The choice of  $W^{g^{}_L}=W^{g^{}_R}=1$ indeed does not exhaust all the possibilities. Particularly, as we inspect the examples given in Ref\cite{Mesaros2011},  models have been directly constructed where the symmetry transformations implementing $g^{}_L$ and $g^{}_R$ anti-commute. Before diving into a thorough comparison of the \km construction with other  constructions, we would like to explore such a possibility in \km construction,
and specifically by understanding the $\Z_2$ topological theories.

Therefore, to construct a model such that the action of the generators $g^{}_L$ and $g^{}_R$ satisfy non-trivial commutation relations and at the same time allowing for the possibility that charges fractionalize, $W^{g^{}_L} \neq W^{g^{}_R}$. Let us therefore consider $K= 2\sigma_x$ and make the following choice
\be
W^{g^{}_L}= 1,\qquad W^{g^{}_R} = \sigma_x.
\ee

The corresponding $\dphib^{g^{}_L}$ and $\dphib^{g^{}_R}$ are then given by
\be
\begin{aligned}
&\dphib^{g^{}_L} = \pi (t_1+ \frac{n^L_1}{2}, t_2+ \frac{n^L_2}{2})^T\\
&\dphib^{g^{}_R} = \pi (\delta_1,\delta_2)^T, \qquad \delta_1+\delta_2 = n^R_2 \pmod{2},
\end{aligned}
\ee
where $n^{L(R)}_{1,2}$ correspond to the identity action $\dphib^\mathbf{e}$ we choose for the group relation $g^2_{L(R)}= 1$, and that for consistency we require also that $n^R_1=n^R_2$.

One could now compare the action of $g^{}_L g^{}_R$ and that of $g^{}_R g^{}_L$.
They now lead to different shifts, which are given by
\be
\begin{aligned}
&\dphib^{g^{}_Lg^{}_R}= \frac{\pi}{2}(n^L_1+ 2t_1+2\delta_1, n^L_2+2t_2 +2\delta_2)^T\\
&\dphib^{g^{}_Rg^{}_L}= \frac{\pi}{2}(n^L_2+ 2t_2+2\delta_1, n^L_1+2t_1+ 2\delta_2)^T
\end{aligned}
\ee
These relations demonstrate the following.
First, that $t_1$ and $t_2$ can now make a difference since they can determine the eigenvalue of $\dphib^{g^{}_L}$ under $W^{g^{}_R}= \sigma_x$. Second, it is clear that the action of $g^{}_L$ and $g^{}_R$ on a fundamental anyon (ie. $(1,0)^T\text{ or }(0,1)^T$) can be anticommuting if $\dphib^{g^{}_L}$ is an eigen-vector of $\sigma_x$ with eigenvalue $-1$. Nevertheless, such a commutativity is most natural when the representation of $g^{}_L$ is in fact projective; otherwise, a linear representation, which has  $n^L_1=n^L_2=0$, would imply that the action of $\dphib^{g^{}_Lg^{}_R}$ on a fundamental anyon produces a factor of $\exp(\pi)$ while that of $\dphib^{g^{}_Lg^{}_R}$ a factor of $\exp(-\pi)$, which are in fact identical.

There is, however, one special situation where fractionalization is not necessary for anti-commutative actions on a fundamental anyon, which is achieved by taking $t_1=1, t_2=0$ and that $\delta_1=\delta_2=1$. In which case, 
\be
\begin{aligned}
&\dphib^{g^{}_Lg^{}_R}= \pi(2, 1)^T\\
&\dphib^{g^{}_Rg^{}_L}= \pi(1, 2)^T
\end{aligned}
\ee
One can see that each fundamental anyon acquires opposite sign under the action of $g^{}_Lg^{}_R$ and $g^{}_Rg^{}_L$. In this case also since $W^{g^{}_R} $ is proportional
to the \km and also its inverse, the edges cannot be trivially gapped.

\section{$SEP(\Z_4\x \Z_4,\Z_2)$ Phases}\label{sec:Z2Z2xZ2}
We now consider the case where we incorporate a global $\Z_2$ symmetry into the topological phases described by the theories defined by a family of eight \kms:
\be\label{eq:Kn1n2n3}
K=\bpm
-2n_1 & 2  & n_2 & 0\\
2 & 0 & 0 & 0\\
-n_2 & 0 & -2n_3 & 2\\
0 & 0 & 2 & 0
\epm,\,
K^{-1}=\frac{1}{4}\bpm
0 & 2 & 0 & 0\\
2 & 2n_1 & 0 & n_2\\
0 & 0 & 0 & 2\\
0 & n_2 & 2 & 2n_3
\epm,
\ee
where $n_1,n_2,n_3\in\{0,1\}$. These \kms all have $|K|=16$, indicating that there are $16$ quasiparticle types in theory defined by each such \km.  If $n_2=0$, it is clear that these \kms turn out to be the direct sum of the $2\x  2$ \kms in Section \ref{sec:Z2xZ2}; hence, we can infer that with $\Z_2$ global symmetry incorporated, the SET phases will be just those already found in $SEP(\Z_2\x\Z_2,\Z_2)$. New phases with nontrivial boundary modes may thus appear only if $n_2$ is turned on, such that the \km is not block-diagonal. We arrange the three integers $n_1$ through $n_3$ into an array $[n_1n_2n_3]$ and use this to denote the eight cases to be studied.

\subsection{Fusion and Gauge Groups  }\label{subsec:gaugeKn1n2n3} %
In this basis of the \kms, a generic quasiparticle $\lv=(l_1,l_2,l_3,l_4)$ has its components $l_1$ and $l_3$ labeling the charges, while $l_2$ and $l_4$ labeling the corresponding fluxes respectively\cite{Hung2012a}. The self statistics is 
\be\label{eq:sStatKn1n2n3}
\frac{\theta_l}{\pi}=l_1l_2+\frac{1}{2}(l^2_2n_1+l_2l_4n_2+l^2_4n_3)+l_3l_4 \pmod 2,
\ee
which obviously can take values in $\{0,1/2,1,3/2\}$. Hence, there are $16$ elementary quasiparticles all told, consistent with $|\det(K_{[010]})|=16$. We would not record all these elementary quasiparticles here but note that they can be obtained by allowing $l_1$ through $l_4$ in Eq. \eqref{eq:sStatKn1n2n3} to be either $0$ or $1$ and grouped by their self statistics. 

The fundamental quasiexcitations are the two charges: $\mathbf{e}_1=(1,0,0,0)$, $\mathbf{e}_2=(0,0,1,0)$, and the two fluxes: $\mathbf{m}_1=(0,1,0,0)$ and $\mathbf{m}_2=(0,0,0,1)$. These four fundamental excitations all have the bosonic self statistics but not trivial mutual statistics with all other quasiparticles, as can be easily checked. But they can fuse to physical bosons. We would like to nail down the general charge vectors 
of bosons in terms of these fundamental excitations, which also allows us to read off the fusion algebra of the quasiparticles in this theory. 

Let $\lv_B=(l_1,l_2,l_3,l_4)^T$ be a generic boson and $\lv'=(l'_1,l'_2,l'_3,l'_4)^T$ an arbitrary quasiparticle, their mutual statistics is 
\be\label{eq:mStatBosonKn1n2n3}
\begin{aligned}
\frac{\theta_{l_Bl'}}{\pi} 
= &l_2l'_1+(l_1+l_2n_1+\frac{l_4n_2}{2})l'_2\\
&+l_4l'_3+(\frac{l_2n_2}{2}+l_3+l_4n_3)l'_4,
\end{aligned}
\ee
which must be $0\pmod{2}$. The terms in the above
equation are grouped as in the second row therein because the free variables are $l'_1$ through $l'_4$, whereas $l_1$ through $l_4$ are constrained such that the mutual statistics is trivial.
Now that $l'_1$ through $l'_4$ are free and independent, the four terms in the second row of Eq. (\ref{eq:mStatBosonKn1n2n3}) must be equal to $0\pmod 2$ individually. We then infer that the most general constraints on $l_1$ through $l_4$ are $l_2=2b$, $l_3=2c-bn_2$, $l_4=2d$, and $l_1=2a-dn_2$, where $a,b,c,d\in\Z$ are free integer parameters. Quite naturally, these constraints are independent of $n_1$ and $n_3$. Thus, the physical bosons of the theory take the following general form.
\be\label{eq:bosonKn1n2n3}
\lv_B=(2a-dn_2,2b,2c-bn_2,2d),\; a,b,c,d\in\Z.
\ee
We can thus identify the following four elementary bosons:
\be\label{eq:eleBosonKn1n2n3}
(2,0,0,0),(0,2,-n_2,0),(0,0,2,0),(-n_2,0,0,2).
\ee
The fusion algebra is generated by the fusion rules of the previously defined four fundamental quasiparticles, namely $\mathbf{e}_1$, $\mathbf{e}_2$, $\mathbf{m}_1$, $\mathbf{m}_2$. Since bosons are considered equivalent to the trivial particle $\mathbf{0}=(0,0,0,0)$  in the fusion algebra, Eq. \eqref{eq:eleBosonKn1n2n3} leads to the following relations:
\be
\begin{aligned}
&\mathbf{e}_1\x  \mathbf{e}_1=\mathbf{0},\quad \mathbf{e}_2\x  \mathbf{e}_2 =\mathbf{0}, \\
&\mathbf{m}_1\x \mathbf{m}_1\x \mathbf{(e}_2)^{n_2}=\mathbf{0},\quad \mathbf{m}_2\x  \mathbf{m}_2\x \mathbf{(e}_1)^{n_2}=\mathbf{0},
\end{aligned}
\ee
where the exponent is formal, meaning that $\mathbf{(e}_i)^{0}=\mathbf{0}$ and $\mathbf{(e}_i)^{1}=\mathbf{e}_i$, $i=1,2$. It is straightforward to check that the fusion algebra $\F_{[n_1n_2n_3]}$ of the $16$ quasiparticles respecting the above relations turn out to be
\be\label{eq:fAlgKn1n2n3}
\F_{[n_1n_2n_3]}=\left\{
\begin{array}{ll}
\Z_4\x  \Z_4,& n_2=1\vspace{0.5em}\\
(\Z_2\x \Z_2)\x (\Z_2\x \Z_2),& n_2=0
\end{array}
\right..
\ee

These two fusion groups can also be verified by the projective representation $\{W^\mathbf{e},\dphib^\mathbf{e}\}$ of the identity of whichever global symmetry to be incorporated, as we now show. Since this identity must preserve any boson up to a $2\pi$ shift, namely, $\l_B^I\phi_I\rightarrow\l_B^I\phi_I +l_B^I\dd\phi^e_I=l^I_B\phi_I\pmod{2\pi}$, we immediately have $W^\mathbf{e}=\mathds{1}_4$ and the following constraint on $\dphib^\mathbf{e}$
\begin{align*}\label{eq:dphieKn1n2n3}
l_B^I\dd\phi^e_I=&2a\dd\phi^e_1+b(2\dd\phi^e_4-n_2\dd\phi^e_1) \numberthis\\
&+ c(2\dd\phi^e_2-n_2\dd\phi^e_3)+2d\dd\phi^e_3=0\pmod{2\pi},
\end{align*}
where $\lv_B^T$ takes the general form in Eq. (\ref{eq:bosonKn1n2n3}), and the terms are grouped by the free integer parameters $a$ through $d$. As such, each term in Eq. \eqref{eq:dphieKn1n2n3} should be $0\pmod{2\pi}$. Clearly, $\dd\phi^e_1$ and $\dd\phi^e_3$ can always be either $0$ or $\pi$, and their value determines the possible values of $\dd\phi^e_4$ and $\dd\phi^e_2$ respectively. It is therefore not hard to write the allowed $\dphib^\mathbf{e}$ in a compact form as follows.
\be\label{eq:dphieKn1n2n3}
W^\mathbf{e}=1,\quad \dphib^\mathbf{e}=\pi\bpm
t_1\\
t_2+\frac{t_3}{2}n_2\\
t_3\\
t_4+\frac{t_1}{2}n_2
\epm,\,t_{i=1,\dots,4}\in\{0,1\},
\ee 
which readily generate additively ($\hspace{-0.7em}\mod 2\pi$) the fusion group $\Z_4\x \Z_4$ if $n_2=1$ and the group $(\Z_2\x \Z_2)\x (\Z_2\x \Z_2)$ otherwise, as those in Eq. \eqref{eq:WgRotK010}. Again for $G_s$ generated by a single generator, the possible ``gauge group'' $N_g$ involved in extending $G_s$ is generated by a $\dphib^\mathbf{e}$ with one specific choice of $t_1$ through $t_4$. There are only two possibilities:  
\be
N_g=\left\{\begin{array}{ll}
\Z_4,\quad & t_1n_2\neq 0 \text{ or } t_3n_2\neq 0\vspace{0.5em}\\
\Z_2,\quad & \text{otherwise}
\end{array}
\right..
\ee
\subsection{Case with $[n_1n_2n_3]=[0n_20]$}\label{subsec:K010}

Seen from Eq. \eqref{eq:dphieKn1n2n3}, $n_2$ dictates whether the \km $K_{[n_1n_2n_3]}$ has two decoupled blocks and thus the form of the fusion group. Since $n_1$ and $n_3$ play no role in the fusion group, let us set them zero, i.e., we have $[n_1n_2n_3]=[0n_20]=[n_2]$. The \km in Eq. \eqref{eq:Kn1n2n3} becomes
\be\label{eq:K010}
K_{[n_2]}=\bpm
0 & 2  & -n_2 & 0\\
2 & 0 & 0 & 0\\
-n_2 & 0 & 0 & 2\\
0 & 0 & 2 & 0
\epm,\,
K_{[n_2]}^{-1}=\frac{1}{4}\bpm
0 & 2 & 0 & 0\\
2 & 0 & 0 & n_2\\
0 & 0 & 0 & 2\\
0 & n_2 & 2 & 0
\epm.
\ee
The $GL(4,\Z)$ transformations that preserve $K_{[010]}$ are the matrices as follows
\begin{align*}\label{eq:XmatrixK010}
&X_{\alpha}=\pm\bpm
\mathds{1}_2 & \alpha \boldsymbol\delta_{21}\\
-\alpha\boldsymbol\delta_{21} & \mathds{1}_2
\epm
,
X_{\beta}=\pm\bpm
\mathds{1}_2 & (\beta-1)\boldsymbol\delta_{21}\\
\beta\boldsymbol\delta_{21} & -\mathds{1}_2
\epm, \\
&X_{\gamma}=\pm\bpm
\gamma\boldsymbol\delta_{21} & \mathds{1}_2 \\
\mathds{1}_2 & -\gamma\boldsymbol\delta_{21}
\epm
,
X_{\lambda}=\pm\bpm
(\lambda-1)\boldsymbol\delta_{21} & \mathds{1}_2 \\
-\mathds{1}_2 & \lambda\boldsymbol\delta_{21}
\epm,\numberthis
\end{align*}
where $\mathds{1}_2$, and $\boldsymbol\delta_{21}$ are respectively the $2\x  2$ identify matrix, and the $2\x  2$ matrix $(\begin{smallmatrix}0&0\\1&0 \end{smallmatrix})$, and $\alpha,\beta,\gamma,\lambda\in\Z$ parameterize an infinite family of these $X$ matrices.

According to Eqs. \eqref{eq:bosonKn1n2n3} and \eqref{eq:dphieKn1n2n3}, we immediately see that in absence of global symmetry, all edge modes can be gapped by condensing either of the following sets of bosons,
\begin{subequations}\label{eq:indBosonNoSymmKn1n2n3}
\begin{eqnarray}
&\mathbf{A_1}&:=\{(0,2b,2c-n_2b,0)|\,b,c\in\Z\},\label{eq:indAbosonNoSymmKn1n2n3}\\
&\mathbf{A_2}&:=\{(2a,0,2c,0)|\,a,c\in\Z\},\label{eq:indBbosonNoSymmKn1n2n3}\\
&\mathbf{A_3}&:=\{(2a-dn_2,0,0,2d)|\,a,d\in\Z\},\label{eq:indCbosonNoSymmKn1n2n3}
\end{eqnarray}
or any set in the following two infinite one-parameter families of sets $\mathbf{B}_{p\in\mathbb{Q}}$ and $\mathbf{B}_{q\in\mathbb{Q}}$ .
\begin{align*}
\mathbf{B}_p :=&\Big \{(-2cp,2b,2c-bn_2,2bp)\Big| \numberthis\label{eq:indBpbosonNoSymmKn1n2n3}\\
& p\in\mathbb{Q},\; b,c,bp\in\Z ,\;  n_2 \frac{bp}{2}-cp\in\Z \Big \},\\
\mathbf{B}_q :=&\Big \{(-2dq,2b,2bq-bn_2,2d)\Big|\\
& q\in\mathbb{Q},\; b,d,bq\in\Z ,\; n_2 \frac{d}{2}-dq\in\Z \Big \}. \numberthis\label{eq:indBqbosonNoSymmKn1n2n3}
\end{align*}
\end{subequations}
Note that if $n_2=0$, the system is only a stack of two copies of the toric code model that is studied in our first example; hence, the above sets in Eq. \eqref{eq:indBosonNoSymmKn1n2n3} of independent, condensable bosons will recombine to merely four sets, each of which consists of one of the four combinations of the independent bosons respectively in the two toric code model with $\Z_2$ symmetry. 
\subsubsection{Representations of the $\Z_2$ global symmetry}
\label{subsubsec:K010Z2SGrep}
Similar to previous examples, for $\Z_2$ global symmetry to be incorporated, we look for (projective) representations $\{W^g,\dphib^g\}_{g\in\Z_2}$of the $\Z_2$ global symmetry group that transforms the fundamental fields but may allow certain independent Higgs terms. We should first demand that for all $g\in\Z_2$, $(W^g)^2=\mathds{1}_4$ and $(W^g)^T K_{[n_2]}W^g=K$. The latter condition guides us to find the correct $W^g$ matrices from the $X$ matrices in Eq. \eqref{eq:XmatrixK010}; hence, we obtain $W^g=\pm\mathds{1}_4,\, X_{\beta},\, X_{\gamma}$. We are interested in inequivalent $W^g$ transformations, and  since $W^g=X_{\beta}$ and $W^g=X_{\gamma}$ are are related by a $GL(4,\Z)$ transformation preserving the \km, as $X_{\lambda}^{-1}X_{\gamma}X_{\lambda}=-X_{\beta}$, they
are in fact equivalent and will not be considered separately. Moreover, for any value of $\gamma$, one can always apply a $GL(4,\Z)$ transformation by certain $X$ matrix in Eq. \eqref{eq:XmatrixK010} that preserves the \km, while keeping the form of $\dphib^\mathbf{e}$ in \eqref{eq:dphieKn1n2n3} up to redefinition of the parameters $t_1$ through $t_4$, to set $\gamma=0$ in $X_{\gamma}$. Thus, we conclude with the inequivalent $W^g$ transformations
\be\label{eq:Z2WgK010}
W^g=\pm\mathds{1}_4,\, W^g=X_{\gamma=0}=\pm\mathds{1}_2\otimes\sigma_x,
\ee
where $\sigma_x$ is the usual Pauli matrix and $\otimes$ the usual matrix tensor product. Note that the matrices with a $+$ sign and a $-$ sign in the front are not equivalent to each other under the transformation in Eq. (\ref{eq:gaugeWgDphig}).
 
Before we proceed to nail down the corresponding $\dphib^g$, let us remark on the behavior of $W^g=\pm\mathds{1}_2\otimes\sigma_x$. The action of $W^g$ on a quasiparticle $\lv$ is given by $(W^g)^T\lv$, and to manifest the physics we let $\lv=(e_1,m_1,e_2,m_2)$, where $e_1$ and $m_1$ ($e_2$ and $m_2$) are respectively the charge and flux associated with the first $\Z_2$ (second $\Z_2$) gauge group of the total $\Z_z\x\Z_2$ gauge group. Then for $W^g=\mathds{1}_2 \otimes\sigma_x$ we have 
\be\label{eq:WgRotK010}
(W^g)^T\lv=(\pm\mathds{1}_2\otimes\sigma_x)^T\lv=\pm \bpm e_2\\ m_2 \\ e_1\\ m_1\epm,
\ee  
which signifies a non-local exchange of the two types of dyons, $(e_1,m_1,0,0)$ and $(0,0,e_2,m_2) $ respectively of the two $\Z_2$ sectors of the gauge group. Such a non-local exchange transformation by the global symmetry is evidently beyond the scope of symmetry fractionalization, as also reported in Ref\cite{Mesaros2011}.
Note that this exchange transformation exists for any choice of $[n_1n_2n_3]$, even if $n_2=0$.

We now solve for $\dphib^g$. Since in any extension of $\Z_2$ by $\Z_2\x\Z_2$, the latter exists as a normal subgroup; hence, the group compatibility conditions demands that
\be\label{eq:dphigK010}
(\mathds{1}+W^g)\dphib^g=\dphib^\mathbf{e},
\ee
for any $\dphib^\mathbf{e}$ in Eq. \eqref{eq:dphieKn1n2n3}. We solve the above equation for $\dphib^g$ in different cases of $W^g$.

\textbf{(i)} $W^g=-\mathds{1}_4$. Equation \eqref{eq:dphigK010} has the unique, inequivalent solution $\dphib^g=\mathbf{0}$ and $t_1=t_2=t_3=t_4=0$ must be set in $\dphib^\mathbf{e}$. Since $\cos(l^I\dd\phi^g_I)$ is invariant under $W^g=-1$, the global symmetry $\Z_2$ does not transform the quasiparticles at all, implying that the edge modes can be completely gapped out, resulting in a boundary-trivial phase that is identical with the phase without the global symmetry.

\textbf{(ii)} $W^g=\mathds{1}_4$. The solution of Equation \eqref{eq:dphigK010} clearly is 
\be\label{eq:dphigWgIdK010}
\dphib^g=\pi\bpm p_1\\p_2\\p_3\\p_4\epm+\frac{1}{2}\dphib^\mathbf{e},\, p_{i=1,\dots,4}\in\{0,1\}.
\ee  

This nontrivial shift vector in general prevents the edge modes from being fully gapped, as it forbids any of the sets of independent variables in Eq. \eqref{eq:indBosonNoSymmKn1n2n3}. Special cases do exist, e.g., $t_1=t_3=0$ in $\dphib^g$ would allow the entire set $\mathbf{A_2}$ to condense, resulting in a boundary-trivial phase. Nevertheless, a thorough study of all boundary-nontrivial phases and their quasi-group structure in this case turns out to be rather complicated because we lack of a convenient and systematic algorithm for computing the new sets of independent, condensable bosons in a stacking of many phases for large-size \kms. While we are not able to unveil the full quasi-group structure of the phases in this case, we do have a partial result to summarize as follows but  fill the details in Appendix \ref{appd:phaseZ2Z2xZ2}. 

A study of how the independent bosons in Eq. \eqref{eq:indBosonNoSymmKn1n2n3} transform by the shift vector in Eq. \eqref{eq:dphigWgIdK010} show that the relevant parameters in Eq. \eqref{eq:dphigWgIdK010} are $t_1,\,T_2= t_2-n_2p_3,\, t_3$, and $T_4=t_4-n_2p_1$, where new parameters $T_2$ and $T_4$ are defined in terms of the old ones. As such, our experience tells us that we can label all possible phases by the values of the string $[t_1 T_2 t_3 T_4]$, leading to 16 phases. Tabulated in Appendix \ref{appd:phaseZ2Z2xZ2}, 12 out of these 16 phases actually have fully gapped edge state without symmetry breaking. There are four  edge-nontrivial phases remaining in Eq. \eqref{eq:unGapK010Wg1} with non-trivial edge:
\begin{align*}\label{eq:ungapPz2z2xZ2}
[t_1T_2t_3T_4]=&[1010]\\
               &[1011]\\
               &[1110]\\
               &[1111].\numberthis
\end{align*}
We have not explored the quasi-group relations between these four phases, which gets cumbersome as larger \kms are involved. This should be worth a future attempt.

\textbf{(iii)} $W^g=\pm\mathds{1}_2\otimes\sigma_x$.
In this case, one can apply the equivalence transformation in Eq. (\ref{eq:gaugeWgDphig}) first to turn arbitrary $\dphib^g$ into a common simpler form by removing any redundancy. It is not hard to show that by choosing $X=\mathds{1}$ in Eq. (\ref{eq:gaugeWgDphig}), for any $\dphib^g$, one can always find a shift $\Dphib$ to eliminate the first two component of the $\dphib^g$, without affecting $W^g$. As such, one can assume that in general, $\dphib^g= (0,0,x,y)^T$, where $x$ and $y$ are  to be solved. The equation above now becomes
$(\pm x,\pm y,x,y)^T=\dphib^\mathbf{e}$, which is soluble only when $t_1=\pm t_3$ in $\dphib^\mathbf{e}$, leading to
\be
\dphib^g=\pm \pi\bpm 0\\ 0\\ t_1\\ t_2+\frac{t_1}{2}n_2\epm,
\ee
with constraints $t_1=\pm t_3$ and $t_4=\pm t_2$ on $\dphib^\mathbf{e}$ enforced.

Interestingly, however, since this $\dphib^g$ does not yield any nontrivial shift to the boson variables in the set $\mathbf{A_2}$ in Eq. \eqref{eq:indBbosonNoSymmKn1n2n3}, as $l^I_{\mathbf{A_2}}\dd\phi^g_I\equiv \pm2\pi ct_1=0\pmod{2\pi},\,\forall \lv_\mathbf{A_2}\in\mathbf{A_2}$, one can gap out all the edge modes by condensing the independent Higgs terms constructed from the bosons in set $\mathbf{A_2}$ as follows.
\be
\sum_{a,c}C_{a,c}\int\dd t\dd x \big\{\cos[2(a\phi^e_1+c\phi^e_2)] +\cos2(a\phi^e_2 +c\phi^e_1)]\big \},
\ee
where $\phi^e_{1(2)}$ are respectively the electric edge modes associated with respectively the left and the first and the second $\Z_2$ factors of the $\Z_2\x \Z_2$ gauge group.
Therefore, despite a nontrivial exchange of the two quasiparticle types under $W^g=\mathds{\pm 1}_2\otimes\sigma_x$ and even symmetry fractionalization due to the nontrivial $\dphib^g$, the corresponding phase remains boundary-trivial.

\section{Beyond central extension}\label{sec:beyond}

Our examples in the previous two sections demonstrate some novel
features in the transformation properties of anyons when
one relaxes a crucial requirement imposed in Ref\cite{Essin2012}, namely
that the symmetry operators transforming the anyons have to be local.
In the previous two sections, we have provided several examples where the exchange 
of anyons, a glaringly non-local transformation, can give rise to more exotic phases,
some of which for example has already been reported in Ref\cite{Mesaros2011}. 

There is another important class of phases which also generally
involve non-local transformation of the anyons. Reiterating Ref\cite{Essin2012}, restrictions to
local transformations has led to a classification of phases via different allowed projective representations consistent with the fusion rules. That, in other words, is equivalent to a
classification of different \emph{central} extensions of
the global symmetry group $G_s$ by an Abelian \emph{gauge group} $G_g$
that is taken to be the fusion algebra $G_g= \F$ of the topological
phase on top of which global symmetry is built\cite{Essin2012}. 
The restriction to central extensions has been raised to include more general
group extensions\cite{Hung2012a}. In this case, the global symmetry group becomes
the quotient group $G_s= G/G_g$, and the gauge group is the normal 
subgroup of a bigger group $G$. Different phases correspond to different choice
of the \emph{total} group $G$ for given $G_s$ and $G_g$. In this construction, $G_g$ is no longer the
center of the group $G$, and so one does not expect the group action of $G_g$ and 
$G_s$ to commute. 
From the previous sections therefore, it almost immediately follows that such
group actions necessarily involve exchange of anyons. In fact, the examples
in the previous sections can be understood within this framework of general
group extension. 

In this section we would like to make such a construction within the \km framework 
more explicit, and illustrate these principles using a particular set of examples, where the
total group $G$ is chosen to be one of the dihedral groups $D_{N}$ for some 
odd $N$.

\subsection{Step 1: Obtaining a linear representation of $G$}
The virtue of identifying a total group $G$ in the classification of phases
with symmetry is that any non-trivial or non-linear transformation under the action of the
global symmetry group $G_s$ can be reduced to a simple linear group action in a suitable
 $G$. Here, we will focus our attention on realizing $G= D_N$ for $N$ odd via $K$-matrices.

In $D_N$ one can specify each group element by a pair $(A, a)$, where $A= \pm$, and $a=\{0,1,\cdots, N-1\}$. Group multiplication between two such pairs is given by
\be
(A,a) \cdot (B,b) = (A.B, ( A b + a)_{\textrm{mod} N}).
\ee

A representation of each group element $(A,a)$ is given by
$(W^{(A)}, \dphib^{a})$, where $W^{(+)} = \mathds{1}$ and $W^{(-)} = W^{(-)}$ for some non-trivial $W^{(-)}$ that keep the $K$- matrix invariant, and that $(W^{(-)})^2 = \mathds{1}$. On the other hand, 
$\dphib^{a} = \frac{2\pi a}{N} \dphib^{(-)}$, where $\dphib^{(-)}$ is an eigen-vector of
$W^{(-)}$ with eigenvalue $-1$. The aggregate action of $(A,a)\cdot(B,b)$ is then given by
\be
\begin{aligned}
&W^{(A)}(W^{(B)}{\phi} + \dphib^{b} )+  \dphib^{a} \\
&= W^{(A)} W^{(B)}\phi + W^{(A)}\dphib^b + \dphib^a.
\end{aligned}
\ee
Since $\dphib^{b} $ is an eigenvector of $W^{(A)}$ with eigenvalue $A$, and that
components of $\phi$ are defined only up to multiples of $2\pi $, 
we conclude that the above representation is a faithful representation of $D_N$. 

In the special case where $K= N\sigma_x$ for example, $W^{(-)}$ can be chosen to be
$W^{(-)}= \sigma_x$, and $\dphib^{(-)\,\,T} = (1,-1) $.

\subsection{Step 2: Identifying the normal subgroup with the \emph{gauge} group}
Having constructed a linear representation, we would then have to identify a normal
subgroup $G_g$ of the total group $G$ such that the group action of $G_g$ is
taken to be \emph{unphysical}. In other words, $G_g$ is taken as some kind of
\emph{gauge} group that does not have any visible effect on any physical,
or \emph{gauge invariant}, excitations.  Therefore, admissible 
$G_g$ is strongly restricted by the fusion group $\F$. In fact they 
 are embedded inside $\F$. In other words, the normal subgroup 
$G_g$ can only be chosen whose group action on physical bosons in a \km model
is trivial.  i.e.
\be
\lv_B^T (W^{g}\phi + \dphib^{g}) = \lv_B^T \phi \pmod{2\pi}.
\ee

Now returning to our dihedral group $D_N$. Suppose we would
like to pick the $\Z_N$ normal subgroup as our gauge group. This $Z_N$ subgroup
consist of pairs $(A=+, a)$, where the first component is $+$. The group action
is then given by $\{W^{(+)}= \mathds{1}, \dphib^{a}\}$. This can be admitted
as a gauge group only if $\lv_B^T\dphib^{a}=0\pmod{2\pi}$.  This already suggests
that $Z_N$ must be at least a subgroup of the fusion group.
For example $K= N\sigma_x$, where $\lv_B^T= N(n_1,n_2)$, for any
$n_i \in \Z$, and $\dphib^{a} = 2\pi a/N (1,-1)^T$  , 
indeed we have
$\lv_B^T \dphib^a =0 \pmod{2\pi}$,
and therefore we are allowed to take $\Z_N$ to be a gauge group.

\subsection{Step 3: Implementing the global symmetry group}
The global symmetry $G_s = G/G_g$. In this case where $G= D_N$
and $G_g= \Z_N$, $G_s= \Z_2$. The group elements of $G_s$ are
the cosets of $G$ w.r.t $G_g$.
The identity element of $G_s$ is the coset which is in fact spanned
by the nomral subgroup
$G_g$ itself. In this case therefore, it comprises all the pairs $(+, a)$. Other cosets
are generated by the normal subgroup by left multiplication  
$g_g \times g$, $g_g\in G_g$, and $g \in G$. We note that right multiplication
would yield identical cosets for normal subgroups $G_g$.
The other non-trivial coset corresponding to the non-trivial element in $G_s= \Z_2$
is the set of pairs $(-, a)$. 

Now, the final step is to pick any representative in one of the non-trivial coset, whose
group action is now interpreted as that of the global symmetry group. It automatically
acts non-linearly on the anyons. Its action is closed as a group, up to group action of $G_g$
 which is now so aligned with the fusion algebra $\F$ that physical bosons transform 
trivially. 
For our example at hand, we can take the generator of $G_s=\Z_2$ for $K= N\sigma_x$
to be $\{\sigma_x,\tfrac{2\pi a}{N} (1,-1)^T\}$, for any $a \in \{0,1,\cdots N-1\}$. 
Note that the shift $2\pi a/N (1,-1)$ on any boson lead to shifts proportional
to $2\pi$. Therefore we need only to worry about $W^{(\pm)}$. Given
that $W^{(-)}= \sigma_x$, it immediately reduces to a situation we have
encountered already in subsection \ref{subsec:return}, where
not a single edge mode can be gapped as we continue stacking; hence, in this case, the quasi-group of the phases is $\Z$.

We note that the idea of central extensions work in precisely the same
way, except that the group action is restricted to be commutative. Here we
demonstrate how a non-trivial group extension can be implemented within
the framework of \km construction.

\section{Comparison with other works}\label{sec:Comp}
Endowing topological phases with symmetry is a novel
and important question that has been a subject of much interest recently.
In the previous sections, we have provided yet another construction
of these phases based on \km. Among the scenarios we have
studied, various have already been discussed in the literature. We would therefore
like to make a comparison with known results.

\subsection{Comparison with Ref\cite{Essin2012}}\label{subsec:withHermele}
To begin with, we comment on the relationship of our work with that  in Ref\cite{Essin2012}.
In Ref\cite{Essin2012}, the main targets are Abelian topological phases endowed with global
symmetries whose action is localized near the vicinity of the anyons excited 
in the system. In those cases, it is demonstrated that the anyons can
transform under projective representations of the symmetry group concerned.
These projective representations are consistent with fusion rules: namely that the 
identity element must act trivially on any physical bosonic excitations, 
even if the bosonic state
is a composite of fused anyons. This constrains the possible projective
representations allowed for individual anyons. There are limited
choices of how the identity element of the symmetry group can act on any 
anyon.
In our explicit construction via \kms,
it is clear that the requirements on $\dphib^\mathbf{e}$  coincide with the discussion of
allowed action of the identity element. Among all the specific cases we studied, of
$\mathcal{G}=\Z_N$ \emph{gauge theories} and their twisted versions such as the double semion model for $N=2$,
every single consistent choice of projective representations as dictated in Ref\cite{Essin2012}, which
are classified by $H^2(G_s, \F)$ is realized in our constructions.

\subsection{A comparison with Ref\cite{Hung2012a}}\label{subsec:withWen}
It is also of interest to compare our work with Ref\cite{Hung2012a}.
In Ref\cite{Hung2012a} it is proposed that a systematic construction of
topological phases with symmetry is to consider topological terms
of SPT theories
with symmetry group $G$, whose normal subgroup $N$ is subsequently gauged.
Such a theory should describe a topological phase with global symmetry given by the
quotient group $G_s= G/N$. Specific examples where $G= \Z_4$ and separately $\Z_2\times\Z_2$ are considered, in which $N$ and $G_s$ are both 
$\Z_2$ in each case.
Therefore, these two possible $G$'s correspond to different (central) extensions of 
the global symmetry group. Moreover, for each choice of $G$ there
are several choices of topological terms, classified by $H^3(G,U(1))$. They 
led to many different possible phases. One distinguishing feature between
these different phases constructed is the braiding of excitations around \emph{magnetic}
charges of the global symmetry group -- by magnetic charges, they really correspond to
multi-valued field configurations with branch cuts that end at a branch point.   
In the \km construction however, all field configuration is single-valued, and these
extra braiding statistics are invisible to us. If we ignore them, then there is a one-to-one
correspondence between the phases we constructed and the phases studied there. 
In the  case where $G= \Z_4$, there are four phases constructed in Ref\cite{Hung2012a}, which
is parameterized by a topological term with coefficient $m$ which can take values in $\{0,1,2,3\}$. The correspondence with our construction is as follows:
\be
\begin{aligned}
m=0 \qquad & K10 \\
m=2 \qquad& K11 \\
m=1 \qquad& S01 \\
m=3 \qquad& S10
\end{aligned}
\ee
On the other hand, when $G=\Z_2\times \Z_2$ there are eight phases with three independent 
topological terms, each parameterized by a coefficient $n_i = \{0,1\},$ for $i=\{1,2,3\}$.
It is demonstrated there\cite{Hung2012a} that fractionalization occurs if and only if $n_2$ is non-vanishing.
There are four phases therefore that admit fractionalization, and again each of them directly
corresponds to our construction:
\be
\begin{aligned}
(010) \qquad & K01 \\
(110) \qquad & K01 \\
(011) \qquad & S11\\
(111) \qquad & S11
\end{aligned}
\ee

We note that in the above, two phases are mapped to the same \km phase because as emphasized already these phases differ only if magnetic charges of the global symmetry is visible, which they are not in our \km construction. It is perhaps also surprising that here all the phases have trivial edge excitation!

We finally note that the general proposal of Ref\cite{Hung2012a} generalizes the classification of phases 
via central extension to a general group extension of the global symmetry group $G_s$, although explicit examples considered there lie within the framework of central extension.
In this work we provide first such examples of a general group extension realizing the proposal in Ref\cite{Hung2012a}.

\subsection{Comparison with  Ref\cite{Mesaros2011}}\label{subsec:withYR}

Finally, we would like to comment on the relationship of
our work with that of Ref\cite{Mesaros2011}. 
In Ref\cite{Mesaros2011}, a (lattice) topological gauge theory as defined in Ref\cite{Dijkgraaf1989,Dijkgraaf1990} is taken as a starting point, whose
gauge group $G$ is chosen to be a direct product of $\mathcal{G}$ and $G_s$. The action amplitude
of the theory is characterized by difference choice of ``topological terms'' $\nu$ which are
group cocycles in $H^3(G,U(1))$. 
The $G_s$ part is then \emph{ungauged}, by restricting field configurations to be 
pure gauge. i.e.  For each field configuration where degrees of freedom sit on the links of the lattice, each of
which labeled by a pair $(g_g,g_s)$, where $ g_g\in \mathcal{G}$ and $g_s\in G_s$,  each $g_s$ at a particular link in the collection of degrees of freedom can always be written as $g_s= s_i s_j^{-1}$,
where $s_{i,j} \in G_s$ and $i,j$ label the vertices connected by the link concerned. In otherwords, each set of link variables  $\{g_s\}$ can be replaced by a set of vertex variables $\{s_i\}$. $\mathcal{G}$ however remains gauged, supplying the long-range entanglement needed
in a topological phase. 
It is observed in various explicit examples that the pure electric excitations always
transform linearly under $G_s$, and that by picking different $\nu$  magnetic or dyonic excitations of $\mathcal{G}$ can transform non-linearly. In particular, anyons can transform in different non-trivial
representations of the global symmetry group $G_s$, including the projective representations as discussed in Ref\cite{Essin2012}, but not restricted to them. 

We specifically wish to comment on two of our examples which are motivated by observations in Ref\cite{Mesaros2011}.  Before that, one should note the role played by $\mathcal{G}$ in these topological gauge theory constructions.
In particular one should be cautious and observe that the residual gauge group $\mathcal{G}$ is not to be confused with the fusion algebra $\F$ between \emph{all} the anyons. As already
emphasized in our overview of the paper, $\mathcal{G}$ is the ``deconfined'' gauge group, and it is (subgroups of) $\F$ that is often being identified as the \emph{gauge} group in Ref\cite{Essin2012}, which we have denoted $N_g$ throughout most of our paper. 
It is however as expected that in these models, $\mathcal{G}$ forms the sub-fusion algebra involving only the pure electric excitations. 
Without going into technical details, we can identify the \km description that corresponds to $\mathcal{G}= \Z_2$ and $G_s= \Z_2\times \Z_2$, which is discussed in section \ref{subsec:Z2Z2SGZ2G}. We found examples where the action of the generators of the two $\Z_2$ symmetries anticommute.  In section \ref{sec:Z2Z2xZ2}, we also found examples corresponding to $\mathcal{G}= \Z_2\times \Z_2$ and $G_s= \Z_2$, where the 
group action of $G_s$ is manifestly non-local and exchange the gauge electric and magnetic charges between the two $\Z_2$ gauge groups. Both of these cases are considered in Ref\cite{Mesaros2011} and where such novel transformations are also observed. As already mentioned above, while these novel transformations are absent among pure electric charges in Ref\cite{Mesaros2011}, they are ubiquitous in the \km construction, which
naturally provides the flexibility to incorporate non-linear transformations on any excitations,
as long as they are consistent with the full fusion algebra. For this reason, our specific choice of the group action on the electric and magnetic charges may not generally coincide with
the examples in Ref\cite{Mesaros2011}. We believe that the distinction between \emph{electric} and \emph{magnetic} excitations there is an artifact of the topological gauge theory
construction. Given the direct connection between $\mathcal{G}$ and the fusion algebra of purely electric excitations, one realizes that to achieve non-trivial transformations also among pure electric charges in the framework set forth in Ref\cite{Mesaros2011},
one is compelled to consider topological gauge theories where $G$ is taken to be a non-trivial extension of $G_s$ by $\mathcal{G}$ other than the direct product that is being considered\cite{Hung2012a}.  

\section{Discussions and Outlook}\label{sec:Disc}

In this paper, we have been studying intrinsically topological phases endowed with a global symmetry -- the symmetry enriched phases as we dubbed -- aiming at their classification and edge-state properties, by means of the celebrated \km formulation of effective theories of Abelian topological order. While methodologically we extends the application in Ref\cite{Lu2012a,Levin2012a} of \kms in classifying SPT phases to LRE phases with symmetry, we systematically adapt and integrate several principles   imposed in Ref\cite{Essin2012,Lu2012a,Mesaros2011,Hung2012a,Bais2009}, particularly of how a global symmetry may transform the anyons in a topological phase. These principles and the \km method guide us to constructing examples of symmetry enriched phases, along with clarifying a few important conceptual questions, particularly the roles of
various different groups play in classifying different phases. As noted in Ref\cite{Essin2012} it is the fusion group $\F$ of the anyons under consideration that constrains the action of a global symmetry $G_s$, in a way such that the identity of $G_s$ acts trivially on any physical bosons although it may transform any individual anyon exotically, in which case the anyons may undergo local symmetry charge fractionalization and perhaps accompanied by non-local transformations such as anynon exchange. The ``gauge group'' $N_g$ involved in extending $G_s$ is a subgroup of $\F$ that is projected (as the kernel of the projection map) into the identity of $G_s$ and thus preserves the bosons, which indicates the existence of a larger group $G$ that contains $N_g$ as its normal subgroup and $G_s$ is its quotient group $G/N_g=G_s$. Therefore, two different $G_s$ actions  do not commute in general, nor does the action of $G_s$ and that of $N_g$, as shown in some of our examples.

The \km approach offers a convenient way of analyzing the relations among the symmetry enriched phases by stacking the phases, in the sense of arranging the \kms respectively characterizing the phases into a direct sum and the corresponding $G_s$ representations in a direct sum in the same order. In the examples we have shown, the various symmetry enriched phases for a given $\F$ and $G_s$ constitutes a quasi-group structure. In particular in the case with $\F=\Z_2\x\Z_2$ and $G_s=\Z_2$, as explained in Section \ref{subsec:withWen}, the phases in the corresponding quasi-group are identified with the phases under the same setting in Ref\cite{Hung2012a}. To emphasize the prominent role $\F$ plays in these symmetry enriched phases we label the phases accordingly as $SEP(\F,G_s)$. This notion not only covers the symmetry enriched LRE phases but also embraces the SPT phases: the fermionic SPT phases with a given symmetry $G_s$ comprise $SEP(\Z_2,G_s)$ because the fusion group of fermions is $\Z_2$, whereas the bosonic SPT phases all fall into $SEP(\Z_1,G_s)$ because bosons have trivial fusion group $\Z_1$.

Most of the examples we constructed are inspired by Ref\cite{Mesaros2011}, but there
are important differences that should be noted in our construction and discussion. First, we have carefully defined the notion of ``gauge group''. In particular, similar to Ref\cite{Essin2012}, it is
what we denoted $N_g$ that is pertinent: whereas in Ref\cite{Mesaros2011}, the term ``gauge group'' refers exclusively to what we have denoted $\mathcal{G}$ 
Second, the constructions in Ref\cite{Messaros2012} gives rise only to flux fractionalization; however, the \km method treats charge and flux on an equal footing, naturally allowing charge, flux, and dyons to fractionalize simultaneously. 

The \km method has another virtue: it enables us to study the fate of the edge modes explicitly, obtaining the condition when a phase may have gapless edge modes protected by the symmetry. Seen in the examples we constructed, symmetry charge fractionalization or more exotic transformations of the anyons under global symmetry in a LRE phase is neither a sufficient nor a necessary condition for the phase to possess non-trivial edge states.  Although we do not know if these phases that have trivial edges,  despite displaying exotic transformations under the action of the global symmetry, may still be adiabatically connected to an LRE phase without any symmetry, as far as the edge property is concerned, in the quasi-group of all phases in a given $SEP(\F,G_s)$, we may treat those phases having a trivial edge on an equal footing with the phase with the same fusion group but without the symmetry, as if they are projected into the identity of the quasigroup. 

Our first example, i.e., $SEP(\Z_2\x\Z_2,\Z_2)$, is also partly discussed in Ref\cite{Essin2012}, which already exemplifies the important role the fusion group of anyons plays. We realized every phase 
that appears in the classification in Ref\cite{Essin2012}. Our construction however also involves phases that do not appear in \cite{Essin2012}, when we allow for non-local group actions that exchange anyons. Furthermore, in Ref\cite{Essin2012}, the classification of symmetry enriched phases is equivalent to the classification of the central extensions of $G_s$ by $N_g$; however, our examples
also include a non-central extension of $G_s=\Z_3$ by $\Z_2$ to the dihedral group $D_3$,
as anticipated in Ref\cite{Hung2012a}.

Inspired by \cite{Bais2009}, having observed that the various groups involved in characterizing a symmetry enriched phase are related by $\F\supset G\supset N_g$ and $G/N_g=G_s$, and that $N_g$ acts trivially but $G_s$ acts nontrivially on the condensed edge modes, we are encouraged to redraw our picture of symmetry enriched phases as an example of the Hopf symmetry breaking, first proposed and phrased in Ref\cite{Bais2002,Bais2003,Bais2009} to account for anyon condensations, generalizing Landau's symmetry breaking. This new paradigm of generalized symmetry breaking may become most suitable to cope with the non-Abelian anyons endowed with a symmetry. We shall report our detailed studies elsewhere\cite{Hung}. 

Let us close with a discussion of interesting questions that should be more thoroughly addressed in the future. 
We now describe them briefly below.  

\textbf{1.} While we have a detailed analysis of $SEP(\Z_2\x\Z_2,\Z_2)$ that probably exhausts all the phases in the class, our treatment of other examples requires further analysis. Particularly it would be of interest to explore whether a qusi-group structure can be generally defined. 
At the moment it appears that the order of any such quasi-group in $\Z_N$ gauge theories 
grows at least linearly in $N$, which makes an analysis very cumbersome quickly. A more efficient method is necessary for a thorough understanding. 

\textbf{2.} As far as clarifying a group structure of $\Z_N$ gauge theories are concerned, there is another specific question to be addressed. In our first example, we have seen two different models with topological order, i.e., the double semion model and the toric code model, which share the same fusion group. When the same symmetry group is incorporated, they together lead to a set of symmetry enriched phases belonging to the same quasi-group: in particular the $\Z_2$ symmetric double semion model acts as a generator. These two models are actually described by the set of \kms $(\begin{smallmatrix}0&2\\2&2n\end{smallmatrix})$ with a single paramter $n=\pm 1$, which defines the double semion model when $n=1$ and the toric code model when $n=0$. In more general cases, a class of different models of topological order can be specified by a multi-parameter \km. For instance, the \km in Eq. \eqref{eq:ZnKmatrix} characterizes respectively $N$ models respectively for the $N$ values of the parameter $l=0,\dots,N-1$. These models have different fusion groups as in Eq. \eqref{eq:fAlgebraZn}. Ultimately, this parameter $l$ labels the $N$ $3$-cocycles in the cohomology group $H^3(\Z_n,U(1))$ that classifies the corresponding $N$ models, we are not able to answer at this moment the question whether the symmetry enriched phases characterized by respectively the fusion groups in Eq. \eqref{eq:fAlgebraZn} with the same symmetry group $G_s$ would belong to the same quasi-group in a nontrivial way, as opposed to simple direct product of the quasi-groups characterized respectively by the $N$ fusion groups and $G_s$. We are not able to answer this question in general either and hence leave it for future exploration.

\textbf{3.} We have considered only discrete gauge groups and unitary symmetries in this paper. It is of interest to construct more cases with continuous symmetry groups $G_s$ and also those involving time reversal.

\textbf{4.} Having observed non-local transformations of quasiparticles under $G_s$, e.g., the dyon exchange discussed below Eq. \eqref{eq:WgRotK010}, and since non-locality is rather intrinsic to non-Abelian anyons, we look forward to extending our studies to the interplay between non-Abelian topological order and global symmetry. Unfortunately, this is beyond the reach of the \km formalism and thus begs for new approaches.

\textbf{5.} A recent paper by Vishwarnath and Senthil\cite{Vishwanath2012} found that some symmetry enriched
topological phases in $2+1$ dimensions can only exist as the boundary of some SPT phase
in $3+1$ dimensions. We have realized some new phases based on general group extensions using the \km. It would be interesting to understand if the \km or strictly $2+1$ models can exhaust all the phases based purely on consideration of group extensions, or whether some
extra phases are again only realizable as boundaries of higher dimensional non-trivial phases.

\textbf{6.} In the last stage of preparing this manuscript, we noticed a very recent paper by Levin\cite{Levin} that studies the conditions that allow for gapless edge states in a pure Abelian, non-chiral topological order without any global symmetry.  It turns out that non-trivial edges can appear and that they are protected by the quasiparticle braiding statistics in the bulk, instead of by any symmetry.  One such example is the $\nu=2/3$ fractional Quanthum Hall system. The topological phases studied in our paper however have fully gapped edges in the absence of symmetry.  It is of interest to extend our investigation to incorporating global symmetry in these novel phases discussed in Ref\cite{Levin}.    

As we finish our paper, we were brought to the attention of the work of Lu and Vishwanath \cite{Yuan-MingLu2013} which  contains also substantial discussion on $\Z_2$ gauge theories and the doubled semion model enriched by $\Z_2$ symmetries. The number of phases they have obtained in cases restricted to local on-site symmetry action is exactly twice as ours. The extra phases there can be obtained by stacking each of our phase, namely $\{T00, T10,T11,S00,S10,S01,S11\}$ with a non-trivial $\Z_2$ SPT phase, leading altogether to 6 distinct $T$ phases and 8 $S$ phases.  It would be of interest to understand possible extra phases also in the other constructions we have in the current paper by stacking them with SPT phases.

\begin{acknowledgements}
We thank Juven Wang and Dr. Peng Gao for helpful discussions. In particular, we thank Yuan-Ming Lu and  Ashvin Vishwanath for sharing with us their results.  LYH is supported by the Croucher Fellowship. YW acknowledges Prof. Guifre Vidal from whom he got to know the concept of symmetry enriched phases. YW also appreciates Prof. Seigo Tarucha and Prof. Rod Van Meter.
\end{acknowledgements}

\appendix

\section{Phases in $SEP(\Z_4\x\Z_4,\Z_2)$ with $W^g=\mathds{1}$}\label{appd:phaseZ2Z2xZ2} In this appendix we explain how one may arrive at Eq. (\ref{eq:ungapPz2z2xZ2}). Given the much experience one may gained with looking for condensable bosons in the other examples in this paper, here we shall be as brief as we can. Taking the scalar product of the independent bosons in the sets $\mathbf{A_1}, \mathbf{A_2}, \mathbf{A_3}, \mathbf{B}_p,\text{ and } \mathbf{B}_q$ in Eq. (\ref{eq:indBosonNoSymmKn1n2n3}) with the $\dphib^g$ in Eq. (\ref{eq:dphigWgIdK010}), we can find all the relevant terms involving the parameters in $\dphib^g$, which are the terms that are not immediately equal to $0\pmod{2\pi}$, and are tabulated as follows.
\be\label{eq:relTermK010Wg1}
\begin{array}{c|c}\hline
\mathbf{A_1} & bT_2 + ct_3 \\\hline
\mathbf{A_2} & at_1 + ct_3 \\\hline
\mathbf{A_3} & at_1 + dT_4 \\\hline
\mathbf{B}_p & (\tfrac{bn_2p}{2}-cp)t_1+bT_2+ct_3+bpT_4-c'pt_1+c't_3  \\\hline
\mathbf{B}_q & \begin{aligned}(\tfrac{dn_2}{2}-dq)t_1+bT_2+bqt_3+dT_4\\
+(\tfrac{d'n_2}{2}-d'q)t_1+d'T_4\end{aligned}  \\\hline
\end{array}
\ee   
where $T_2=t_2-n_2p_3$ and $T_4=t_4-n_2p_1$. Note that the coefficients in the last two rows of in the above equation must meet the constraints in Eqs. (\ref{eq:indBpbosonNoSymmKn1n2n3}) and (\ref{eq:indBqbosonNoSymmKn1n2n3}). Now that the only relevant parameters are $t_1$, $T_2$, $t_3$, and $T_4$, we can use a string $[t_1T_2t_3T_4]$ to labels all possible phases in this case; hence there are 16 of them. Since the sets $\mathbf{A_1}$ to $\mathbf{A_3}$ are apparently much simpler than the infinite families of sets, we first see which among the 16 phases can have their edge modes fully gapped by condensing the bosons in these simpler sets.  It is immediate from the first three rows in Eq. \eqref{eq:relTermK010Wg1}, as long as any one of the pairs $(T_2,t_3)$, $(t_1,t_3)$, and $(t_1,T_4)$ is $(0,0)$, one can condense the bosons in the set  whose relevant terms are turned off by the corresponding vanishing pair of parameters. Thus, the following phases have trivial edge modes as completely gapped.
\be
\begin{array}{c|l}\hline
[t_1T_2t_3T_4] & \text{Condensable sets}\\\hline
[0000] &\text{Any one} \\\hline
[0001] &\mathbf{A_1} \text{ or } \mathbf{A_2} \\\hline
[0010] &\mathbf{A_3}\\\hline
[0100] &\mathbf{A_2} \text{ or } \mathbf{A_3}\\\hline
[0101] &\mathbf{A_2} \\\hline
[0110] &\mathbf{A_3} \\\hline
[1000] &\mathbf{A_1} \\\hline
[1001] &\mathbf{A_1}\\\hline
\end{array}
\ee        

Now we have 8 phases left. We have two infinite families  of sets at our disposal. Consider the  family $\mathbf{B}_p$ first, if we take $p\in\Z+\tfrac{1}{2}$, i.e., half integers, the constraints $c'p\in\Z$ and $bp\in\Z$ in Eq. (\ref{eq:indBpbosonNoSymmKn1n2n3}) demands that $c',b,c\in 2\Z$; hence, we can let  $b=2j$, $c'=2k$ and assumed $p=1/2$ for simplicity without losing any generality, which renders another constraint $(bpn_2/2-cp)\in\Z$ in Eq. (\ref{eq:indBpbosonNoSymmKn1n2n3})  as $j=2m+c$ with $m\in\Z$. With these in mind, the two relevant terms of $\mathbf{B}_p$ become
\begin{align*}
&(\frac{jn_2}{2}-c)t_1+2jT_2+ct_3+2mT_4+cT_4-k t_1+2kt_3\\
=&\frac{jn_2}{2}t_1 + c(t_3+T_4)\pmod{2},
\end{align*}
where an overall $\pi$ factor is dropped. Then clearly, if $t_1=0$ and $t_3+T_4=2$, the equation above automatically hold, indicating that the edge modes in phases $[0011]$ and $[0111]$ can be completely gapped out by condensing any set $\mathbf{B}_{p\in\Z+1/2}$.

Now let us turn to the family $B_q$.  Since, as argued before in Section \ref{subsec:K010}, $n_2=0$ is equivalent to stacking two copies of toric code model that is studied in our first example, we focus on $n_2=1$ from now on for simplicity. First consider $q=(4k-1)/2$, $k\in\Z$, according to  Eq. (\ref{eq:indBqbosonNoSymmKn1n2n3}), this readily constrains that $b=2j\in 2\Z$ and that $d'/2-d'q=2kd'+d'$.  Hence, the two relevant terms of $\mathbf{B}_q$ in Eq. (\ref{eq:relTermK010Wg1}) become
\begin{align*}
&(\tfrac{d}{2}-d\frac{4k-1}{2})t_1+2jT_2+2j\frac{4k-1}{2}t_3+dT_4\\
&+(2kd'+d')t_1+d'T_4 \\
=&(d+d')(t_1+T_4)-jt_3\pmod 2
\end{align*}
where $n_2=1$ is assumed. Thus, if $t_3=0$ and $t_1+T_4=2$, the two relevant terms will become irrelevant, implying that the edge modes in the phase $[1101]$ can be fully gapped by condensing any $\mathbf{B}_{q\in 2\Z-1/2}$. 

One can verify by similar procedures that condensing any set $\mathbf{B}_q$ with $q\in 2\Z+1/2$, the phase $[1100]$ has a completely gapped edge without breaking the symmetry. We catalog the above results in the following table.
\be
\begin{array}{c|l}
[t_1T_2t_3T_4] & \text{Condensable sets}\\\hline
[0011] &\mathbf{B}_p,\;p\in 2\Z-1/2 \\\hline
[0111] &\mathbf{B}_p,\;p\in 2\Z-1/2 \\\hline
[1100] &\mathbf{B}_q,\;q\in 2\Z+1/2 \\\hline
[1101] &\mathbf{B}_q,\;q\in 2\Z+1/2 \\\hline
\end{array}
\ee
We remark that the second column in above is not meant to be complete, in the sense that other choices of $p$ and/or $q$ may also do the job. But the point is that no set of independent bosons can condense without breaking symmetry to gap the edge modes of the remaining four phases:
\be\label{eq:unGapK010Wg1}
[t_1T_2t_3T_4]=[1010],\;[1011],\;[1110],\;[1111].
\ee

\section{Some useful matrices}
\kms of the form 
\be
K_{[N,l]} = \bpm
0 &N\\
N& 2l
\epm ,
\ee
for $l \in \{0,1,\cdots,N-1\}$ feature frequently in our discussion of
topological phases which descend from deconfined $\mathcal{G}=\Z_N$ gauge theories.

We give a list of $SL(2,\Z)$ matrices $X(N,l)$ that keep $K(N,l)$ invariant.

There are three special cases where there are general solutions of $X$. 
\be\begin{aligned}
&X_{[N,0]} = \sigma_x, \\
&X_{[N,1]} = \bpm 1 &0 \\ -N & -1 \epm \\
&X_{[N,N-1]} = \bpm N-1& N-2 \\-N & 1 - N\epm
\end{aligned}
\ee 
More generally, it can be parametrized as
\be
X_{[N,l]} = \bpm h/k & (h^2 - k^2) s/N \\-N/(k^2 s) & -h/k \epm,
\ee
where $l = h \times k\times s$ for $h,k,s\in \Z$. The parameterization follows
from Euler's parametrization of Pythagorean triples. We note that not all $l$
therefore allow for an $X_{[N,l]} \in SL(2,\Z)$.

\section{From SPT to Topological phases}

As noted first in Ref\cite{Levin2012} and elaborated in Ref\cite{Hung2012,Mesaros2011,Hung2012a}, there
is a close relation between a bosonic SPT phase with symmetry group $G_s$
and a topological phases characterized by a deconfined gauge group $\mathcal{G}$
where $\mathcal{G}=G_s$. The precise relation one can turn an SPT phase with only
short range entanglement into a topological gauge theory with long range entanglement by
introducing an extra set of gauge fields and gauging $G_s$.

This procedure has a direct analog also in the context \km construction. 

Recall that a generic bosonic  SPT phase can be constructed by taking $K= \sigma_x$
as the starting point and then imposing global symmetry by incorporating suitable
Higgs terms that respect the symmetry\cite{Lu2012a}. 
The Chern-Simons Lagrangian is thus
\be
L_{K} = - \frac{1}{4\pi} \epsilon_{\mu\nu\rho} a^I_\mu K_{IJ} \partial_{\nu}a^J_\rho 
\ee
Let us be specific and consider in particular SPT phases with $\Z_N$ symmetry.
In that case, the symmetry transformation is characterized by Ref\cite{Lu2012a}
\be
\{W^g, \dphib^T \} = \{\mathds{1}, 2\pi/N  (1,q)\}.
\ee

This dictates how the anyonic excitations characterized by charge vector $\lv$
created by $b_i = \exp(\ii l^I\phi_I)$ transform. Recall also that the Chern-Simons
construction is the ``dual frame'' description of these bosons\cite{Zee1995}, where
the currents of these bosons are related to the CS gauge fields by $j_\mu = \ii\epsilon_{\mu\nu\rho}/(2\pi) \partial_\nu a_\rho$. (This is a standard normalization. 
See e.g. Ref\cite{Zee1995,Sachdev2012}.)
Therefore we can write down the current of the $\Z_N$ symmetry in terms of $a^I_\mu$,
which is given by
\be
j_\mu = \frac{\ii \epsilon_{\mu\nu\rho}}{2\pi} \partial_{\nu} (a^1_\rho + q a^2_\rho).
\ee

Following the standard procedure, we gauge the $\Z_N$ symmetry by minimally coupling it to 
a gauge field 
\be
L_{\textrm{gauge}} = - j_\mu A^1_\mu,
\ee
where we understand that while $A^1_\mu$ is a $U(1)$ gauge field, we are
preserving only a $Z_N$ subgroup by restricting $A^1_\mu$ to take discrete values $2\pi a/N$, for some integer $0\le a < N$.

In the topological gauge theory, we have conservation of both the electric and magnetic charges.
We should introduce therefore another gauge field $A^2_\mu$ that couples to
magnetic excitations of the ``global turned local'' symmetry. As is already evident in
the discussion in Ref\cite{Hung2012,Mesaros2011,Hung2012a}, the excitation of the gauge fields of the gauged $\Z_N$ is responsible for generating these magnetic configurations. Another way to see that is that in the $L_{\textrm{gauge}}$ term, by an integration by parts $\partial A^1$ becomes
\emph{electric} sources of the $a^{1,2}$ gauge fields, and it is well known  that the electric charges of $a$ correspond to vortex excitations of the bosons $b$ alluded to above 
in the ``direct'' frame. Therefore we introduce the following coupling 
\be
L_{\textrm{magnetic}} = -\frac{N}{4\pi} \epsilon_{\mu\nu\rho}\partial_{\mu}A^1_{\nu} A^2_\rho.
\ee 
The normalization is also dictated by the fact that we expect unit \emph{electric} charge coupled to $A^1$
should gain a phase of $2\pi/N$ when it moves around a unit magnetic charge coupled to $A^2$.
(c.f. discussion in Ref\cite{Hung2012a}). 

The total Lagrangian is then given by
\be
L = L_{K}+L_{\textrm{gauge}} + L_{\textrm{magnetic}}.
\ee
Finally, let us integrate out $a^{1,2}$.  Since this is a quadratic action, this procedure can be most readily done by obtaining their
equations of motion from the total action and then evaluating $L$ on-shell. 
The equations of motion are
\be
da^1 + \ii dA^1=0\,\qquad da^2 + \ii q dA^1 =0.
\ee
We end up with
\be
L = -\frac{\epsilon_{\mu\nu\rho}}{4\pi}  A^I_\mu \mathcal{K}_{IJ} \partial_{\nu}A^J_{\rho},
\ee 
where
\be
\mathcal{K} =\bpm 2 q & N\\ N &0\epm
\ee
which is the expected \km of the topological phase corresponding to
a (deconfined) topological gauge theory with gauged $\Z_N$. 

This procedure can be readily checked for other bosonic SPT phases.
One could readily check that the same procedure works for more general Abelian symmetry groups, such as $\Z_2\times \Z_2$.
We note that our gauging procedure depends on the fact that a $\Z_N$ symmetry can be understood as a subgroup of $U(1)$, which admits a natural gauging procedure. For more general
non-Abelian discrete symmetries we believe an analogous procedure should exist by embedding it in a non-Abelian Lie group.

\bibliographystyle{unsrtnat}
\bibliography{library.bib}

\end{document}